\documentclass[twocolumn, aps, prx, superscriptaddress, longbibliography]{revtex4-1}

\usepackage[utf8]{inputenc}
\usepackage{hyperref}
\usepackage{amsmath, amssymb, amsfonts, mathrsfs, dsfont}
\usepackage{xcolor}
\usepackage{graphicx}
\usepackage{ bbold }
\usepackage{thesis_defs}
\usepackage{feynmp-auto}

\begin{document}

\title{Non-Markovian Dynamics of Open Quantum Systems via Auxiliary Particles with Exact Operator Constraint}

\author{Tim Bode}
\email[E-mail: ]{t.bode@fz-juelich.de}
\affiliation{Physikalisches Institut and Bethe Center for Theoretical Physics, Universit\"at Bonn, Nussallee 12, 53115 Bonn, Germany}
\affiliation{Institute for Quantum Computing Analytics (PGI-12), Forschungszentrum Jülich, 52425 Jülich, Germany}

\author{Michael Kajan}
\affiliation{Physikalisches Institut and Bethe Center for Theoretical Physics, Universit\"at Bonn, Nussallee 12, 53115 Bonn, Germany}

\author{Francisco Meirinhos}
\affiliation{Physikalisches Institut and Bethe Center for Theoretical Physics, Universit\"at Bonn, Nussallee 12, 53115 Bonn, Germany}

\author{Johann Kroha}
\email[Email: ]{kroha@physik.uni-bonn.de}
\affiliation{Physikalisches Institut and Bethe Center for Theoretical Physics, Universit\"at Bonn, Nussallee 12, 53115 Bonn, Germany} 

\begin{abstract}
   We introduce an auxiliary-particle field theory to treat the non-Markovian dynamics of driven-dissipative quantum systems of the Jaynes-Cummings type. It assigns an individual quantum field to each reservoir state and provides an analytic, faithful representation of the coupled system-bath dynamics. We apply the method to a driven-dissipative photon Bose-Einstein condensate (BEC) coupled to a reservoir of dye molecules with electronic and vibronic excitations. The complete phase diagram of this system exhibits a hidden, non-Hermitian phase transition separating temporally oscillating from biexponentially decaying photon density correlations within the BEC. On one hand, this provides a qualitative distinction of the thermal photon BEC from a laser. On the other hand, it shows that one may continuously tune from the BEC to the lasing phase by circumventing a critical point. This auxiliary-particle method is generally applicable to the dynamics of open, non-Markovian quantum systems.
\end{abstract}

\maketitle

\section{Introduction}

Experimental platforms coupling a set of photonic cavity modes to an ensemble of dye molecules, comprised of two-level systems (TLS) of electronic excitations and local vibrational degrees of freedom, are relevant 
for applications ranging from Bose-Einstein condensates (BEC) of photons  
\cite{klaers2010bose, klaers2012statistical, schmitt2014observation,
  hesten2018decondensation, walker2018driven}, exciton polaritons
\cite{keeling2020bose, ramezani2019ultrafast} and plasmonic lattices
\cite{hakala2018bose} to single-photon sources for quantum information
\cite{Clear2020}. Despite general non-equilibrium field theory being available \cite{sieberer2016keldysh}, the stationary states and dynamics of such open, driven-dissipative quantum gases are unexplored to a large extent. Recently, measurements of the second-order coherence in photon BECs \cite{ozturk2020observation, ozturk2019fluctuation} have opened the door to a deeper understanding. In fact, the experiments revealed a hidden, complex structure in the temporal correlations which undergo a non-Hermitian phase transition as the system is driven away from equilibrium, while the spectral mode occupation still follows an equilibrium Bose-Einstein distribution. 

As open quantum systems are driven far from equilibrium, reservoir frequency scales like the spectral substructure~\cite{Schmitt2018} and relaxation rates can no longer be considered fast compared to the photonic cavity loss and tunneling rates. In particular, in multiple coupled cavities~\cite{Kurtscheid2019}, fast Josephson oscillations can approach the time scales of such reservoir processes.
In these cases, quantum coherence and non-Markovian memory effects~\cite{Clear2020, de2017dynamics} are expected to become important.
The theoretical treatment of this non-Markovian regime is complicated
by the vast number of reservoir degrees of freedom and the necessity of describing their full quantum dynamics. Direct numerical time-evolution methods are limited to rather small systems, and the 
extension to larger systems requires truncation of the quantum entanglement between the subsystems~\cite{del2018tensor}. 
On the other hand, approximate techniques based on the Born-Markov approximation and rate equations~\cite{Kirton2013,
Kirton2015, Keeling2016,Radonjic2018,Marthaler2011}, well established for Markovian dynamics even in large systems, inherently cannot capture non-Markovian memory or quantum coherence effects. 
Standard field-theoretic techniques suitable for a large number of degrees of freedom, like the quantum Langevin approach~\cite{lebreuilly2018pseudothermalization}, are hampered by the non-canonical statistics of the pseudospin operators involved in the electronic TLS dynamics, which have, for this reason, been approximated by bosons \cite{Piazza2021}. A recent mean-field model of organic polartions~\cite{Keeling_2022} rests on the assumption of symmetry breaking for the cavity mode and is thus limited to large photon numbers. Few-photon effects such as decondensation cannot be described by this approximation. Cumulant expansions such as employed in~\cite{Keeling_2020} have been shown recently to be difficult to control~\cite{PhysRevResearch.5.033148}.

\begin{figure}[htb!]
    \centering
    \includegraphics[width=\linewidth]{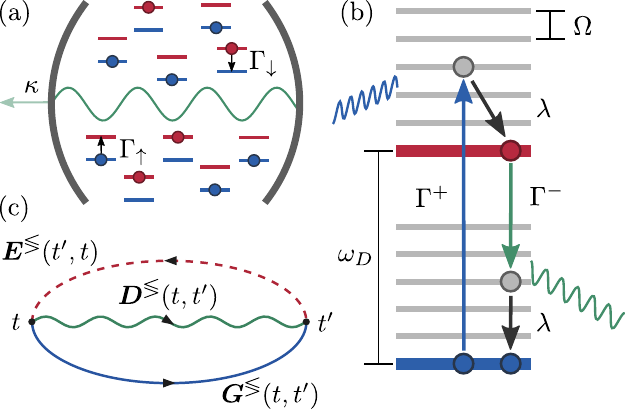}
    \vspace{-0.5cm}
    \caption{(a) Sketch of the dye-filled microcavity. (b) A typical cycle of absorption $\Gamma^+$ and emission $\Gamma^-$ with intermediate vibrational relaxation $\lambda$. (c) Diagram of the Luttinger-Ward functional generating the self-energies~\eqref{eq:selfenergies}.}
    \label{fig:Fig_1}
\end{figure}

In this work, we develop a general field-theoretic technique based on an auxiliary-particle representation, which captures the non-canonical reservoir dynamics exactly, to describe a large class of coherent, light-matter coupled, open quantum systems such as the Holstein-Tavis-Cummings Hamiltonian or arbitrary TLS in a cavity, and generalize it to strong, time-dependent non-equilibrium. In this approach, an individual canonical quantum field is assigned to each matter state (electronic and vibronic molecule excitations), with the constraint that the system occupies exactly one of these states at any instance of time. Because of the canonical commutation relations, the auxiliary-field dynamics are amenable to standard field-theoretic techniques, and the exact confinement to the physical Hilbert space defined by the above-mentioned constraint is performed analytically as long as correlations between different molecules can be neglected. The auxiliary-particle method has been pioneered by Abrikosov~\cite{abrikosov1965electron} and Barnes~\cite{Barnes1977,Barnes1977_II} and is successfully used~\cite{Bickers1987,Coleman1984,kroha1998fermi,kroha2005JPSJ} for strongly correlated electron systems like the Anderson impurity model~\cite{anderson1961localized}. The extension to stationary non-equilibrium has been developed in \cite{Langreth1991,Wingreen1994,hettler1998nonequilibrium,kroha2002nonequilibrium}, and a more recent exposition for correlated electrons in time-dependent non-equilibrium can be found in \cite{Eckstein2010}. 
Generalizing this to quantum optics, we find an exceptional point in the dynamics of the second-order coherence for large losses, and uncover a non-equilibrium decondensation phase transition indicated by critical slowing down. We demonstrate how non-Markovian effects become relevant in the strongly driven-dissipative regime.

\section{Methods}

We consider a general class of Hamiltonians of the form 
\myEq{
    H = H_0 +  H_{\text{ep}} + H_{\text{JC}}, 
}
where
\myEq{
    H_0 = \sum_{k}\omega_{k}^{\phantom{\dagger}} a^{\dagger}_{k}a^{\phantom{\dagger}}_{k} +  \sum_{m=1}^{M} {\omega_D \sigma ^z_m}/{2} + \Omega\, b^{\dagger}_{m}b^{}_{m}
}
describes a set of cavity modes with dispersion $\omega_k$ alongside $M$ molecules comprised of an electronic TLS with splitting $\omega_D$ and a phonon mode $\Omega$ (the generalization to multiple phonon modes is straightforward). The Jaynes-Cummings (JC) coupling in the rotating-wave approximation is as usual given by
\myEq{
    H_{\mathrm{JC}} = g\sum_{k,\,m}(a^{\phantom{\dagger}}_{k}\sigma^+_m + a^{\dagger}_{k}\sigma^{-}_m ).
}
In these expressions, $a^{\dagger}_{k}$, $b^{\dagger}_{m}$ are the creation operators for a photon or a phonon quantum, respectively, and $\sigma^z_m$, $\sigma^{\pm}_m$ are Pauli operators describing the electron dynamics. 
Electron-phonon coupling results from the phononic-oscillator displacement $\hat{x}_m = b_m^{} + b^{\dagger}_m$ depending on the electronic TLS state and reads
\myEq{
    H_{\text{ep}} = \sum_m \Omega\sqrt{S}\,\sigma^z_m\left( b^{}_{m} + b^{\dagger}_{m} \right).
}

\subsection{Auxiliary-Particle Representation}

A single molecule has quantum states $|\sigma,\, n\rangle$, where $\sigma = g,\, e$ refers to the electronic ground and excited state, respectively, and $n$ denotes the vibrational state. For each of these states, we introduce auxiliary or pseudo-boson operators $d^{}_{\sigma,\, n}$, $d^\dagger_{\sigma,\, n}$ with $[d^{}_{\sigma,\, n}, d^\dagger_{\sigma',\, n'}] = \delta_{\sigma\sigma'}\delta_{nn'}$ defined by $d^\dagger_{\sigma,\, n} | \mathrm{vac}\rangle = |\sigma,\, n\rangle$. The molecular operators may then be expressed as
\myEq{\label{eq:aux_bosons}
    b & = \sum_{n=0}^\infty \sum_{\sigma = g, e} \sqrt{n+1}\; d_{\sigma,\, n}^\dagger d_{\sigma,\, n+1}^{\phantom{\dagger}}, \\
    \sigma^+ &= \sum_{n=0}^\infty \excited{n}^\dagger\ground{n}^{\phantom{\dagger}}, \quad
    \sigma^- = \sum_{n=0}^\infty \ground{n}^\dagger\excited{n}^{\phantom{\dagger}} ,
}
where we have dropped the subscripts $m$. This representation is faithful within the Hilbert space spanned by the states $|\sigma,\, n\rangle$ (no product states of these), i.e., under the operator constraint that the total auxiliary particle number obeys
\myEq{
    \hat{Q} = \sum_{n=0}^\infty \sum_{\sigma}d_{\sigma,\, n}^\dagger d_{\sigma,\, n}^{\phantom{\dagger}} = \mathbb{1}.
}
In the frame rotating with the electronic frequency $\omega_D$, the Hamiltonian for $M = 1$ thus reads
\begin{eqnarray}
{H} &=& \sum_{k}\delta_{k}^{\phantom{\dagger}} a^{\dagger}_{k}a^{\phantom{\dagger}}_{k} + \sum_{n=0}^\infty\Big[n\Omega \rbs{  \excited{n}^\dagger \excited{n}^{} +  \ground{n}^\dagger \ground{n}^{} } \nonumber \\
        &+&\Omega\sqrt{S\rbs{n+1}} \rbs{\excited{n+1}^\dagger \excited{n}^{} - \ground{n+1}^\dagger \ground{n}^{} + \mathrm{h.c.}}  \nonumber\\
        &+&  g \sum_{k} \rbs{a_k^\dagger \ground{n}^\dagger \excited{n}^{} + a_k^{} \excited{n}^\dagger \ground{n}^{}} \Big],   \label{eq:Hamiltonian}
\end{eqnarray}
where $\delta_k = \omega_k - \omega_D$ is the resonator detuning. We note in passing that in pseudo-boson representation, the molecular part of this Hamiltonian is quadratic. That is, the polaron transformation diagonalizing this part is straightforward and commutes with the constraint $\hat{Q}=\mathbb{1}$, but transforms the molecule-photon coupling into $\sum_{mn}\gamma_{mn}(a_k^\dagger \ground{m}^\dagger \excited{n}^{} + \mathrm{h.c.})$, where the coupling matrix $\gamma_{mn}$ may be calculated analytically and is proportional to the Franck-Condon integrals. In the following, we will use the non-diagonal representation  Eq.~(\ref{eq:Hamiltonian}) for quantitative evaluations. 

\subsection{Hilbert-Space Projection}  

Since the Hamiltonian (\ref{eq:Hamiltonian}) conserves the operator $\hat{Q}$, any \textit{physical} expectation value of operators acting on the molecular Hilbert space (e.g.\ combinations of $b, b^\dagger$ or $\sigma^{\pm}$ that \textit{annihilate} the empty sector $\hat{Q}=0$) can be projected onto the sector $\hat Q = 1$ by first inserting a factor of $\zeta^{\hat{Q}}$ at the beginning of the Schwinger-Keldysh contour \cite{kamenev2011field}, i.e.\ by replacing the initial density matrix $\hat{\rho}_0$ by $\zeta^{\hat{Q}} \hat{\rho}_0$, where the fugacity is $\zeta = \exp{(-\mu)}$, and the chemical potential $\mu$ can be gauged time-independent. To this end, we write the \textit{canonical} partition function on a Hilbert space $\mathcal{H}_Q$ with any fixed value of the operator constraint as $\operatorname{Tr}_{\mathcal{H}_Q}{\hat{\rho}_0} = Z_{\mathcal{C}}(Q)$. Then the grand-canonical partition function becomes
\myEq{
    Z_{\mathcal{G}} =  \operatorname{Tr}{\zeta^{\hat{Q}}\,\hat{\rho}_0} = Z_{\mathcal{C}}(0) + \zeta Z_{\mathcal{C}}(1) + \mathcal{O}(\zeta^2),
}
where the decisive factor $\zeta^{\hat{Q}}$ weighs each canonical term according to the operator constraint. The expectation value associated with $Z_{\mathcal{G}}$ is
\myEq{
    \E{\hat{X}}_\zeta = Z_{\mathcal{G}}^{-1} \operatorname{Tr}{\hat{X}\,\zeta^{\hat{Q}}\,\hat{\rho}_0}.
}
When $\hat{X}$ annihilates the empty sector $\hat{Q}=0$, its unphysical contribution is removed, and we find the correct average in the physical subspace $\mathcal{H}_1$ as 
\myEq{
    \E{\hat{X}}_{\mathcal{H}_1} = \lim_{\zeta\to 0}\frac{\E{\hat{X}}_\zeta}{\E{\hat{Q}}_\zeta}.
}
It is now instructive to look at two special operator averages at $t_0 = 0$, i.e. 
\begin{align*}
    \begin{split}
        \E{d_{\sigma,n'}^{\phantom{\dagger}} d_{\sigma,n}^{{\dagger}}}_\zeta &\underset{\zeta\to 0}{=} Z^{-1}_{\mathcal{G}} \zeta^0 \operatorname{Tr}_{\,\mathcal{H}_0}\sbs{d_{\sigma,n'}^{\phantom{\dagger}} d_{\sigma,n}^{{\dagger}} \,\hat{\rho}_0} = \mathcal{O}(1), \\
        \E{d_{\sigma,n'}^{{\dagger}} d_{\sigma,n}^{\phantom{\dagger}}}_\zeta &\underset{\zeta\to 0}{=} Z^{-1}_{\mathcal{G}} \zeta^1 \operatorname{Tr}_{\,\mathcal{H}_1}\sbs{d_{\sigma,n'}^{{\dagger}} d_{\sigma,n}^{\phantom{\dagger}} \,\hat{\rho}_0}  = \mathcal{O}(\zeta).       
    \end{split}
\end{align*}
Defining the greater and lesser Green functions
\myEq{
    [\boldsymbol{G}^{>}(t,t')]_{nn'} &= -\ii \E{\ground{n'}^{\phantom{\dagger}}(t) \ground{n}^\dagger(t')}, \\
    [\boldsymbol{G}^{<} (t,t')]_{nn'} &=-\ii \E{\ground{n}^\dagger(t')\ground{n'}^{\phantom{\dagger}}(t)}
}
for the molecular ground states, we conclude that these scale as 
\myEq{
    \boldsymbol{G}^{>} (t,t') &\underset{\zeta\to 0}{=} \mathcal{O}(1),\\
    \boldsymbol{G}^{<}(t, t') &\underset{\zeta\to 0}{=} \mathcal{O}(\zeta)
}
at all times as long as the interactions are treated within a \textit{conserving} approximation as done below. The same holds for the excited-state Green functions 
\myEq{
    [\boldsymbol{E}^{>}(t,t')]_{nn'} &= -\ii\E{\excited{n'}^{\phantom{\dagger}}(t) \excited{n}^\dagger(t')}, \\
    [\boldsymbol{E}^{<}(t,t')]_{nn'} &= -\ii\E{\excited{n}^\dagger(t') \excited{n'}^{\phantom{\dagger}}(t)}.
}
Note that, while the greater functions by themselves do \textit{not} annihilate the ground state, they still occur in Wick-type expansions of physical expectation values such as $\E{\sigma^+(t)\sigma^-(t)}$. For the special case of a TLS without vibrational states (i.e.\ the Jaynes-Cummings model discussed in Appendix~\ref{app:open_JC}), we may calculate physical expectation values such as $\E{\sigma^z(t)}$ by applying the Hilbert-space projection according to 
\begin{align}\label{eq:open_JCM_projected_eqs}
    \begin{split}
        \E{\sigma^+(t')\sigma^-(t)} &\overset{\phantom{t'>t}}{=} \phantom{-}\lim_{\zeta\to 0}\frac{1}{\E{\hat{Q}}_\zeta}\E{d_e^{{\dagger}}(t')d_g^{\phantom{\dagger}}(t') d_g^{{\dagger}}(t)d_e^{\phantom{\dagger}}(t)}_\zeta\\
        &\overset{t'>t}{=} -\lim_{\zeta\to 0} \frac{1}{\zeta}\; \zeta E^<(t, t')G^>(t', t) \\
        &\overset{\phantom{t'>t}}{\equiv} E^<(t, t')G^>(t, t')^*, \\
        \E{\sigma^-(t')\sigma^+(t)} &\overset{\phantom{t'>t}}{=} \phantom{-}\lim_{\zeta\to 0}\frac{1}{\E{\hat{Q}}_\zeta}\E{d_g^{{\dagger}}(t')d_e^{\phantom{\dagger}}(t') d_e^{{\dagger}}(t)d_g^{\phantom{\dagger}}(t)}_\zeta\\
        &\overset{t'>t}{=} -\lim_{\zeta\to 0} \frac{1}{\zeta}\; \zeta G^<(t, t')E^>(t', t) \\
        &\overset{\phantom{t'>t}}{\equiv} G^<(t, t')E^>(t, t')^*,
    \end{split}
\end{align}
where we have dropped the vibrational subscripts ($m, n = 0$) and explicitly pulled out the fugacity. Now, since any product of a greater and a lesser auxiliary-particle Green function is of order $\zeta$, Eq.~\eqref{eq:open_JCM_projected_eqs} recovers the correct result. Note that arbitrary products of time-ordered Pauli matrices can be treated in the same way. For the equal-time average, we hence find the expected result
\myEq{
    \E{\sigma^z(t)} &= \lim_{\zeta\to 0} \frac{1}{\zeta} \rbs{\zeta E^<(t, t)G^>(t, t)^* - \zeta G^<(t, t)E^>(t, t)^*} \\
    &\equiv \ii E^<(t, t) - \ii G^<(t, t),
}
which is plotted in Fig.\ \ref{fig:open_JCM_sigma_z} of Appendix~\ref{app:open_JC}. Note also that we have used the fact that the equal-time greater Green functions follow Eq.~\eqref{eq:aux_boson_equal_time_greater}.

The projection of the molecule dynamics is thus imposed by the simple rule that, in any auxiliary-particle Feynman diagram, only terms of leading order in $\zeta$ are retained. The \textit{projected} Dyson equations then reduce to  
\myEq{\label{eq:proj_Dyson}
    \rbs{\ii \partial_t  - \boldsymbol{h}_G} \boldsymbol{G}^<(t, t') &= \mint{t_0}{t}{\Bar{t}}   \boldsymbol{\Sigma}^>_G(t, \Bar{t})  \boldsymbol{G}^<(\Bar{t}, t') \\ &-  \mint{t_0}{t'}{\Bar{t}}  \boldsymbol{\Sigma}^<_G(t, \Bar{t})  \boldsymbol{G}^>(\Bar{t}, t') , \\
    \rbs{\ii \partial_t - \boldsymbol{h}_G} \boldsymbol{G}^>(t, t') &= \mint{t'}{t}{\Bar{t}}  \boldsymbol{\Sigma}^>_G(t, \Bar{t})  \boldsymbol{G}^>(\Bar{t}, t'),
}
where $\boldsymbol{h}_G$ is the non-interacting part of the Hamiltonian (\ref{eq:Hamiltonian}) in the molecular ground-state sector, and products of bold-faced quantities are understood as matrix products. Analogous equations hold for the excited-state propagators $\boldsymbol{E}^{\lessgtr}(t,t')$. Observe how in the first of Eqs.~\eqref{eq:proj_Dyson}, terms containing two auxiliary-particle lesser functions have vanished, while in the second only terms without lesser functions remain. 

The photon Green functions, defined as 
\myEq{
    [\boldsymbol{D}^{>}(t,t')]_{kk'} &= -\ii\E{a^{}_{k}(t) a^\dagger_{k'}(t')}, \\
    [\boldsymbol{D}^{<}(t,t')]_{kk'} &= -\ii\E{a^\dagger_{k'}(t')a^{}_{k}(t)},
}
do not participate in the $\zeta$-scaling, since they do not operate in the auxiliary-particle space. Thus, their Dyson equations have the usual form,
\myEq{\label{eq:dyson_photons}
     \boldsymbol{D}_0^{-1}\boldsymbol{D}^\lessgtr(t, t')  &= \mint{t_0}{t}{\Bar{t}} \sbs{ \Sigma^>_D(t, \Bar{t}) - \Sigma^<_D(t, \Bar{t})}   \boldsymbol{D}^\lessgtr(\Bar{t}, t') \\
    &-\mint{t_0}{t'}{\Bar{t}}   \Sigma^\lessgtr_D(t, \Bar{t}) \sbs{ \boldsymbol{D}^>(\Bar{t}, t') - \boldsymbol{D}^<(\Bar{t}, t')},
}
where we have defined $\boldsymbol{D}_0^{-1} = \ii \partial_t - \boldsymbol{h}_D$. We also introduce the occupation of the photon ground mode as
\myEq{
    N(t) =  -\mathrm{Im} D_{00}^<(t, t),
}
with steady state $\bar{N} = N(t \to\infty)$.

\subsection{Conserving Approximation}

As mentioned, a conserving approximation is necessary to implement the $\hat{Q}$ conservation. It can be generated from a Luttinger-Ward functional \cite{berges2004introduction, Gasenzer2005, kroha2005JPSJ}. As we show in Appendix~\ref{app:open_JC}, other approximation schemes such as heuristic truncations of the cumulant hierarchy~\cite{Keeling_2020, Radonjic2018} can lead to unphysical effects excluded here by construction. To second order in the coupling $g$, one obtains the analog of the non-crossing approximation~\cite{kroha2005JPSJ}, where the self-energies become 
\myEq{\label{eq:selfenergies}
    \boldsymbol{\Sigma}^{\gtrless}_G(t, t') &= \ii g^2 \boldsymbol{E}^{\gtrless}(t, t')\operatorname{Tr}\sbs{\boldsymbol{D}^{\lessgtr}(t', t)} , \\
    \boldsymbol{\Sigma}_E^{\gtrless}(t, t') &= \ii g^2 \boldsymbol{G}^{\gtrless}(t, t')\operatorname{Tr}\sbs{\boldsymbol{D}^{\gtrless}(t, t')} , \\
    {\Sigma}_D^{\gtrless}(t, t') &= \ii g^{2}\zeta^{-1} \operatorname{Tr}\sbs{\boldsymbol{E}^{\gtrless}(t,t')\boldsymbol{G}^{\lessgtr}(t', t)},  
}
which is depicted diagrammatically in Fig.~\ref{fig:Fig_1} (c). The factor $\zeta^{-1}$ in $\Sigma_D^\gtrless$ stems from the normalization of the physical average. In any self-energy insertion to $\boldsymbol{D}^{\gtrless}$ appearing in an auxiliary-particle diagram, such a normalization is not present. Consequently, such insertions from the same molecule vanish by the projection $\zeta\to 0$. For arbitrary numbers of molecules $M$, the projection onto the physical subspace $\hat Q=1$ is done as described above separately for each molecule $m$ since $\hat{Q}$ is conserved locally. Coherence between different molecules may nevertheless be mediated by the photon modes, which couple to all the molecules.  For multiple molecules, the right-hand side of Eq.~(\ref{eq:dyson_photons}) acquires an additional prefactor $M$, and the photon propagators within the self-energies of any particular molecule $m$ become renormalized by self-energy contributions from all the other molecules $m'\neq m$.

\subsection{Driven-Dissipative Photon-Molecule Dynamics}

While the molecule-bath dynamics are treated fully coherently, we incorporate the external drive and loss by adding Lindblad terms to the master equation for the density matrix,  
\myEq{
\partial_t\rho = \ii[\rho, H] + \sum_{k}\kappa\mathcal{L}[a_k^{}]\rho + \mathcal{L}_{{M}}\rho,
}
where $\mathcal{L}[X]\rho = X\rho X^\dagger - \{X^\dagger X, \rho\}/2$. The cavity loss is $\kappa$ and $\mathcal{L}_{{M}} = \mathcal{L}_{\uparrow,\downarrow} + \mathcal{L}_{\mathrm{relax}}$, where the external drive and radiationless decay of electron excitations are described by (c.f.\ Fig.~\ref{fig:Fig_1}~(b))
\myEq{
\mathcal{L}_{\uparrow,\downarrow} = \sum_{n}\cbs{ \Gamma_{\uparrow} \mathcal{L}[\excited{n}^\dagger \ground{n}^{}] + \Gamma_{\downarrow} \mathcal{L}[\ground{n}^\dagger \excited{n}^{}]},
}
while the process of phonon relaxation due to molecular solvent collisions is captured via 
\myEq{\label{eq:lindblad_relax}
    \mathcal{L}_{\mathrm{relax}} &= \sum_{n,\sigma} {\rbs{n + 1}} \Big(\lambda\rbs{{\bar{n}(\Omega)} + 1}\mathcal{L}[d_{\sigma,\, n}^\dagger d_{\sigma,\, n+1}^{}] \\
    &+ \lambda{\bar{n}(\Omega)}\mathcal{L}[d_{\sigma,\, n+1}^\dagger d_{\sigma,\, n}^{}]\Big),
}
where $\lambda$ is the relaxation rate of the molecular vibrations, and the solvent temperature enters implicitly through the average phonon occupation number $\bar{n}(\Omega)$. To our knowledge, the exact projection method employed here has not, until now, been applied to \textit{open} quantum systems. Therefore, we provide more details on how this generalization is to be performed. Note also that it was not \textit{a priori} clear that such a generalization should be possible. 

The Lindblad operator~\eqref{eq:lindblad_relax} describing the relaxation of the vibrational states can be understood by investigating the structurally equivalent operator
\myEq{
    \frac{\Gamma}{2}\sbs{\rbs{\bar{n} + 1}\mathcal{L}[d_0^\dagger d_1^{}] + \bar{n}\mathcal{L}[d_1^\dagger d_0^{}]}\rho,
}
where $\Gamma$ is now an arbitrary constant. The projection is again performed by removing all terms with too many lesser functions, where one should keep in mind that for auxiliary particles there holds
\begin{align}
    \label{eq:aux_boson_equal_time_greater}
    \boldsymbol{G}^>(t, t)  = \boldsymbol{G}^>(T, 0) &= -\ii\mathds{1}
\end{align}
for all times $T$. We remark again that these properties are \textit{not} an approximation, but exact identities under the projection method. We arrive at
\begin{subequations}\label{eq:supp_aux_boson_proj_vertical_time}
\begin{align}
    \begin{split}\ii\partial_t{\boldsymbol{G}}^<(t, t') 
     &= \boldsymbol{h}_G
    \boldsymbol{G}^<(t, t')\\
    &+\frac{\Gamma \bar{n}}{2}
    \begin{pmatrix}
    {G^>_{11}(t, t)} & 0 \\
    0 & G^>_{00}(t, t)
    \end{pmatrix}  \boldsymbol{G}^<(t, t')\\
    &+ \frac{\Gamma}{2}
    \begin{pmatrix}
    0 & 0 \\
    0 & G^>_{00}(t, t)
    \end{pmatrix}  \boldsymbol{G}^<(t, t'), \end{split}  \\
    \begin{split}\ii\partial_t{\boldsymbol{G}}^>(t, t') &= \boldsymbol{h}_G  \boldsymbol{G}^>(t, t')\\
    &+ \frac{\Gamma \bar{n}}{2}
    \begin{pmatrix}
     {G^>_{11}(t, t)} & 0 \\
    0 & G^>_{00}(t, t)
    \end{pmatrix}  \boldsymbol{G}^>(t, t') \\
    &+ \frac{\Gamma}{2}
    \begin{pmatrix}
     0 & 0 \\
    0 & G^>_{00}(t, t)
    \end{pmatrix}\boldsymbol{G}^>(t, t').\end{split}
\end{align}    
\end{subequations}
These equations evidently possess the correct fugacity scalings. For the equal-time equation of $\boldsymbol{G}^<$, once more we remove all terms with the wrong scaling. This yields
\myEq{
&\ii\dot{\boldsymbol{G}}^<(T, 0) 
    = \left[\boldsymbol{h}_G, 
    \boldsymbol{G}^<(T, 0) \right]\\
    &+\frac{\Gamma}{2}\left\{
    \begin{pmatrix}
    \bar{n}G^>_{11}(T, 0) & 0 \\
    0 & \rbs{\bar{n} + 1}G^>_{00}(T, 0)
    \end{pmatrix},  \boldsymbol{G}^<(T, 0)  \right\}   \\
    &- \Gamma \begin{pmatrix}
    \rbs{\bar{n} + 1}G^<_{11}(T, 0) & 0  \\
    0  & \bar{n} G^<_{00}(T, 0)
    \end{pmatrix} \boldsymbol{G}^>(T, 0),
}
where we have introduced an anti-commutator. Again, this equation has the correct fugacity scaling. As mentioned above, we do not need an equation for the equal-time evolution of $\boldsymbol{G}^>$. Within both the ground- and excited-state manifolds, the Lindblad operators of Eq.~\eqref{eq:lindblad_relax} couple neighboring vibrational states according to a structure which for the self-energy matrices of Eqs.~\eqref{eq:supp_aux_boson_proj_vertical_time} and {$n + 1$} states in total looks like
\myEq{
    \rbs{\bar{n} + 1} &\operatorname{diag}\rbs{0, G^>_{00}, {2} G^>_{11}, ..., {n} G^>_{n-1,n-1}} \\
    + \bar{n} &\operatorname{diag}\rbs{G^>_{11}, {2} G^>_{22},..,  {n} G^>_{nn}, 0}.
}
We still have to consider the external pumping and electronic loss terms  $$\mathcal{L}_{\uparrow, \downarrow} = \sum_{n}\cbs{ \Gamma_\uparrow\mathcal{L}[\excited{n}^\dagger \ground{n}^{}] + \Gamma_\downarrow\mathcal{L}[\ground{n}^\dagger \excited{n}^{}]}.$$ 
The relevant matrices for this read
\myEq{
    \begin{pmatrix}
    \boldsymbol{E}^\lessgtr(t, t')  & \boldsymbol{0} \\
    \boldsymbol{0} & \boldsymbol{G}^\lessgtr(t, t')
    \end{pmatrix}.
}
In terms of these, the contribution to the equations of motion can be written rather compactly as
\begin{align}
    \begin{split}
        &\ii\partial_t\begin{pmatrix}
        {\boldsymbol{E}}^\lessgtr(t, t')  & \boldsymbol{0} \\
        \boldsymbol{0} & {\boldsymbol{G}}^\lessgtr(t, t')
        \end{pmatrix} \\
         &= \frac{\Gamma_\uparrow}{2}
        \begin{pmatrix}
        \boldsymbol{0} & \boldsymbol{0} \\
        \boldsymbol{0} & \operatorname{diag}(\boldsymbol{E}^>(t, t))
        \end{pmatrix}  \begin{pmatrix}
        \boldsymbol{E}^\lessgtr(t, t')  & \boldsymbol{0} \\
        \boldsymbol{0} & \boldsymbol{G}^\lessgtr(t, t')
        \end{pmatrix}.
    \end{split}  
\end{align}
For the forward dynamics, we find
\begin{align}
    \begin{split} 
        &\ii\begin{pmatrix}
        \dot{\boldsymbol{E}}^<(T, 0)  & \boldsymbol{0} \\
        \boldsymbol{0} & \dot{\boldsymbol{G}}^<(T, 0)
        \end{pmatrix} \\
    &= \frac{\Gamma_\uparrow}{2}\left\{
    \begin{pmatrix}
    \boldsymbol{0} & \boldsymbol{0} \\
    \boldsymbol{0} & \operatorname{diag}(\boldsymbol{E}^>(T, 0))
    \end{pmatrix},  \begin{pmatrix}
        \boldsymbol{E}^<(T, 0)  & \boldsymbol{0} \\
        \boldsymbol{0} & \boldsymbol{G}^<(T, 0)
        \end{pmatrix} \right\}\\
    &- \Gamma_\uparrow \begin{pmatrix}
    \operatorname{diag}(\boldsymbol{G}^<(T, 0)) & \boldsymbol{0}  \\
    \boldsymbol{0}  & \boldsymbol{0}     \end{pmatrix} \begin{pmatrix}
        \boldsymbol{E}^>(T, 0)  & \boldsymbol{0} \\
        \boldsymbol{0} & \boldsymbol{G}^>(T, 0)
        \end{pmatrix} .\end{split}
\end{align}
The equations for the electronic loss $\Gamma_\downarrow$ follow analogously. The functional form of the total external pumping reads
\myEq{\label{eq:S_gauss}
    \Gamma_\uparrow(t) &= \Gamma_\uparrow\rbs{1 + A \exp{\cbs{-\frac{\rbs{t - t_P}^2}{2\sigma_P^2}}}},
}
where $t_P$ varies with the attainment of the steady state, and $A = 10^{-2}$, $\lambda\sigma_P = 1/2$.
 \begin{figure}[htb!]
	\centering
	\includegraphics[width=0.9\linewidth]{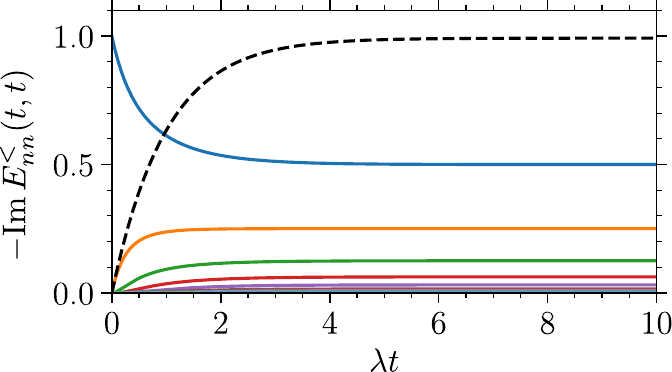}
	\caption{Relaxation of the vibrational degrees of freedom in the electronic excited state. The dashed line shows $-\operatorname{Im}\sum_{n=0}^{p_{\text{max}}} n E^<_{nn}(t, t)\to \bar{n}$. The parameters are $g=0$, i.e. we have decoupled the excited states from the rest of the system, $\omega_D/\lambda = 2.0$, $\Omega/\lambda = 0.1$, $S=0.5$, $\bar{n} = 1.0$, $\Gamma_\uparrow = \Gamma_\downarrow = 0$. The initial conditions is $\ii E^<_{00}(0, 0) = 1$ with all other lesser functions being zero. The number of vibrational states considered is 10, which is equivalent to a maximal number of phonons $p_{\text{max}} = 9$. The results were calculated with a step size of $10\lambda\cdot 2^{-9}$ for a number of $2^9$ steps.}
\label{fig:thermal_phonons}
\end{figure}
The initial conditions are $N(t) = 0$, with $D^>_{00}(0,0)$ following from the commutator, $G^<_{mn}(0,0)=-\ii\delta_{00}$ and $E^<_{mn}(0,0)=0$. The greater functions $G^>_{mn}(0,0) = E^>_{mn}(0,0)=-\ii\delta_{mn}$ are fixed by the projection.

Before concluding this section, consider Fig.\ \ref{fig:thermal_phonons} which shows the relaxation of the vibrational degrees of freedom in the electronic excited state induced by the vibrational relaxation in Eq.\ \eqref{eq:lindblad_relax}. The dashed line indicates the statistical average $-\operatorname{Im}\sum_n nE^<_{nn}(t, t)$ which correctly approaches $\bar{n}$. We therefore conclude that our master equation is capable of describing the phonon thermalization adequately.

\section{Results and Discussion}

\begin{figure*}[htb!]
    \centering
    \includegraphics[width=\linewidth]{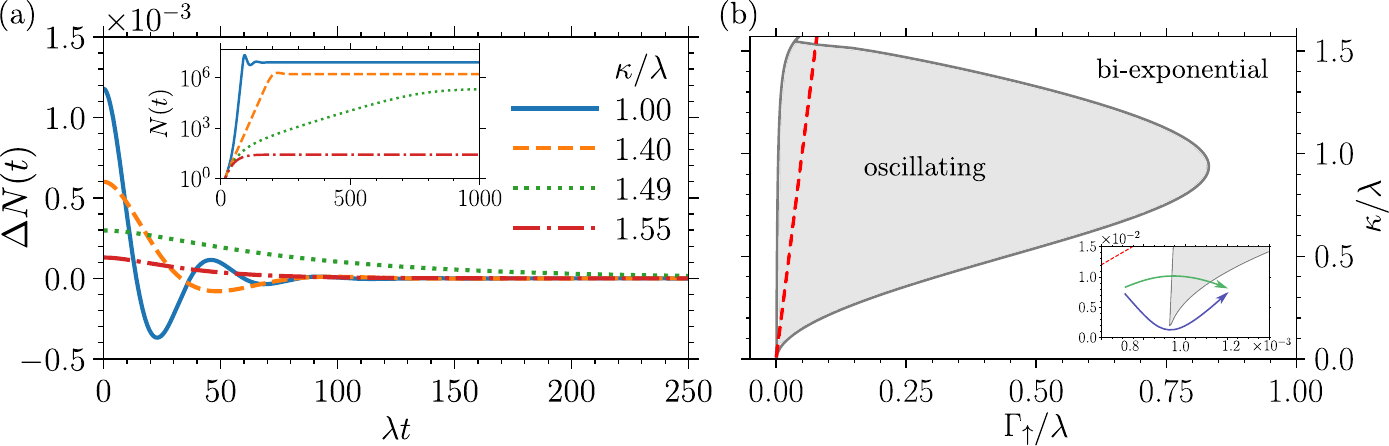}
    \caption{(a) Photon relaxation $\Delta N(t) = N(t) - \bar{N}$ for initial conditions $N(0)=0$, $\ii{G}^<_{00}(0, 0) = 1$, and $M=10^9$, $g/\lambda = 4.5\cdot 10^{-5}$, $\delta_0/\lambda = -1.00$, $\Omega/\lambda = 1.00$, $S=1$, $\bar{n} = 0.25$, $\kappa/\Gamma_\uparrow = 20$, $\Gamma_\downarrow/\lambda = 1.25\cdot 10^{-3}$, $\lambda \tau_{\text{mem}} = 4.0$, and time step $\lambda {\Delta}t = 2^{-4}$. For low solvent temperatures encoded by $\bar{n}=0.25$, phonon truncation at $n_{\text{max}}=4$ is appropriate. Main panel: $t_0$ is chosen at the response peak for each curve. (b) Global phase diagram classifying $\Delta N(t)$, obtained from the rate equations~\eqref{eq:photons_rate_eq}. The boundary is marked by exceptional points in the relaxation spectrum. Red, dashed line: $\kappa/\Gamma_\uparrow = 20$. Inset: avoiding the transition by circumventing the critical point of the oscillatory phase.} 
    \label{fig:Fig_3}
\end{figure*}

\begin{figure}[htb!]
	\centering
	\includegraphics[width=0.95\linewidth]{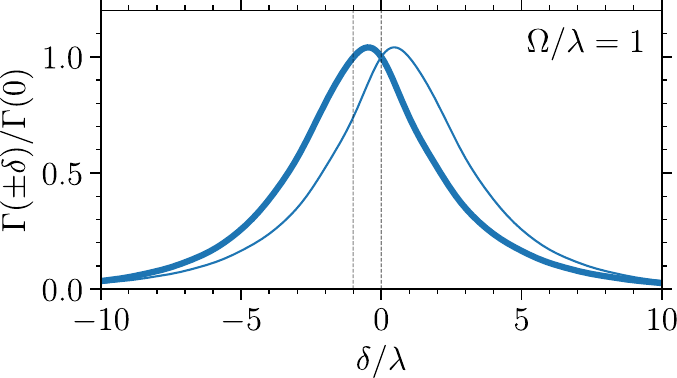}
	\caption{Steady-state emission and absorption spectra $\Gamma^\pm(\delta)$ as calculated from the steady-state photon self-energy for initial conditions $\ii\operatorname{Tr}\boldsymbol{G}^<(0, 0) = \ii\operatorname{Tr}\boldsymbol{E}^<(0, 0) = 1/2$, parameters $S = 1.00$, $\bar{n} = 0.25$, $\Gamma_\uparrow = \Gamma_\downarrow = 0$, vibrational-state truncation at $n=4$, and $\lambda \Gamma(0) = g^2 \cdot 0.798$. The thick line corresponds to the emission coefficient $\Gamma(-\delta)$. Simulations calculated for $2^{11}$ steps up to a final time of $\lambda T_{\text{max}} = 16$ with memory time $\lambda\tau_{\text{mem}} = 4.0$ (defined in Appendix~\ref{app:mem_trunc}).}
\label{fig:Fig_4}
\end{figure}

We solve the Kadanoff-Baym Eqs.~(\ref{eq:proj_Dyson}, ~\ref{eq:dyson_photons}) self-consistently with Eq.~\eqref{eq:selfenergies} using a multi-step predictor-corrector method \cite{Bock2016, Stan2009} developed earlier to facilitate adaptive two-time evolution \cite{Meirinhos2022}. We truncate the time integrals at the memory time $\tau_{\text{mem}}$, which is set by the inverse, dissipative relaxation rates $\kappa$, $\lambda$ (Appendix~\ref{app:mem_trunc}). This allows for an efficient yet accurate simulation. For small system sizes, we find quantitative agreement with quasi-exact numerical time evolution~\cite{del2018tensor}. 
As a further consistency check, in the Markovian limit (phonon relaxation rate $\lambda$ faster than any other scale in the system) our approach reduces to the previously studied semi-classical rate equations~\cite{Kirton2013}, with effective absorption and emission coefficients 
\myEq{
    \Gamma^\pm_k = \Gamma(\pm\delta_k) \equiv -2\operatorname{Re}K(\pm\delta_k),
}
as derived in Appendix~\ref{app:rate_eqs}. $\Gamma(\pm\delta_k)$ encodes the molecular absorption and emission spectra shown in Fig.\ \ref{fig:Fig_4} as a function of $\delta \equiv\delta_k$. Around the zero-phonon line, $\delta = 0$, these spectra satisfy the Kennard-Stepanov relation at the temperature fixed implicitly via $\bar{n}(\Omega)$ (see parameter values in Fig. 3). For the numerical evaluations, we consider a low solvent temperature such that $\bar{n}(\Omega)=0.25$ which justifies truncating the phonon occupation numbers at $n\leq n_{\text{max}}=4$.

Previous experiments \cite{ozturk2020observation} revealed that driven-dissipative photon condensates possess hidden, non-trivial dynamics in the second-order coherence
\myEq{
g^{(2)}(t) = \lim_{s\to\infty}\frac{\E{N(t + s)N(s)}}{\E{N(s)}^2},
}
despite the spectrally resolved photon number following an equilibrium Bose-Einstein distribution. The $g^{(2)}(t)$ oscillation frequency and decay rates as functions of $\Gamma_\uparrow$ and $\kappa$, i.e.\ of the distance from true equilibrium, show a non-Hermitian phase transition marked by an exceptional point \cite{ozturk2020observation}. Deviations from this picture, obtained on the basis of the rate equations~\cite{Kirton2013, ozturk2019fluctuation}, are to be expected when the system is strongly out of equilibrium. Our formalism is ideally suited for studying such non-Markovian effects.

We start the time evolution with an empty single-mode cavity filled with an ensemble of molecules in the ground state and drive the system into a steady state by a constant optical pumping $\Gamma_{\uparrow}$
(inset of Fig.\ \ref{fig:Fig_3}(a)). Once stationarity is reached, a short Gaussian pulse is added (Eq.~\eqref{eq:S_gauss}) to trigger a response of the photon field as the system is slightly displaced from the steady state. Via quantum regression \cite{carmichael2013statistical}, the ensuing relaxation is then equivalent to the spontaneous intensity fluctuations described by $g^{(2)}(t)$.
As shown in Fig.\ \ref{fig:Fig_3}(a), the photon relaxation changes qualitatively from oscillatory to bi-exponential as a function of $\kappa$ (observe that $\kappa/\Gamma_\uparrow$ is kept fixed) \cite{ozturk2020observation,ozturk2019fluctuation}. This behavior is generic and independent of the specific parameters chosen. 
For each set of parameters, the transition between the two behaviors is characterized by an exceptional point where the eigenvalues of the linearized regression dynamics \cite{ozturk2019fluctuation} coalesce and switch from a pair of complex  frequencies $\pm\ii\omega - \tau^{-1}$ to two real relaxation rates $\tau^{-1}$, $\tau'^{\,-1}>\tau^{-1}$ \cite{heiss2012physics}. We extract these parameters by fitting the sum of two complex exponentials to the numerical results Appendix~\ref{app:fit}. 

\begin{figure}[htb!]
    \centering
    \includegraphics[width=\linewidth]{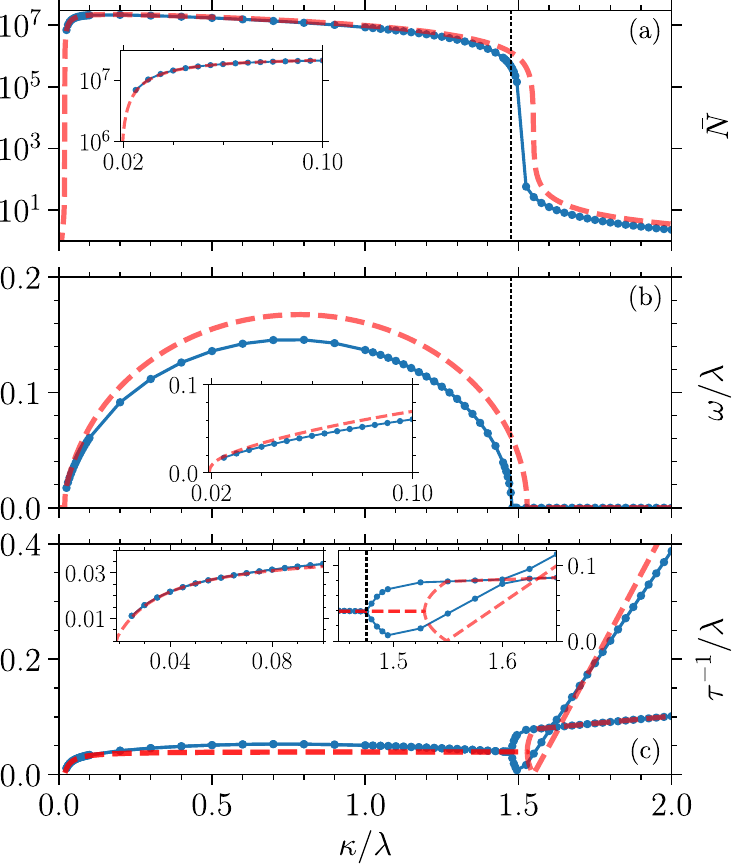}
    \caption{Steady-state photon number $\bar{N}$, and relaxation frequency $\omega$ and rate $\tau^{-1}$ as functions of the cavity loss $\kappa$ for $\kappa/\Gamma_{\uparrow} = \text{const}. = 20 $. Blue, solid curves: full quantum dynamics, \hbox{Eqs.~\eqref{eq:proj_Dyson}-\eqref{eq:selfenergies};} 
    red, dashed curves: rate-equation model (Appendix~\ref{app:rate_eqs}). $\lambda\Gamma_0^- = 0.796 g^2$, $\Gamma_0^+/\Gamma_0^- = 0.741$ (obtained from Fig.~\ref{fig:Fig_4}). The vertical line marks the exceptional point.}\label{fig:Fig_5}
\end{figure}

We find that the phase boundary between both behaviors retraces qualitatively the one obtained from the rate equation model in Appendix~\ref{app:rate_eqs}. Therefore, in Fig.~\ref{fig:Fig_3}(b) we show the global dynamical phase diagram of the photon relaxation as a function of the drive $\Gamma_\uparrow$ and dissipation $\kappa$ as obtained from the rate equations, and compare in Fig. ~\ref{fig:Fig_5} the full quantum dynamics, Eqs.~\eqref{eq:proj_Dyson}-\eqref{eq:selfenergies}, with the rate-equation results along the red, dashed line shown in Fig.~\ref{fig:Fig_3} (b). 
Evidently, the rate-equation model agrees with the full quantum dynamics only near equilibrium ($\kappa/\lambda\ll 1$, $\Gamma_{\uparrow}\ll 1$), while for strong drive and dissipation deviations occur, indicating a breakdown of the Markov assumption inherent to the rate-equation approach: the system becomes non-Markovian in this stronger sense, as opposed to merely a frequency-dependent photo-molecular coupling \cite{Schmitt2018}. 
As seen from Fig.~\ref{fig:Fig_5} (a), the steady-state occupation $\bar{N}$ collapses beyond a critical loss rate $\kappa$ in spite of constant $\kappa/\Gamma_\uparrow$~\cite{hesten2018decondensation}. 
This seemingly counterintuitive effect is due to 
the loss of coherence with increasing $\kappa$ and $\Gamma_{\uparrow}$.
This decondensation is accompanied by critical slowing down, i.e., a diverging relaxation time, $\tau^{-1}\to 0$ (Fig.~\ref{fig:Fig_5} (c)), a clear sign of a non-equilibrium phase transition , while the fast relaxation rate $\tau'$ remains finite. Since $\tau^{-1}\to 0$ is only possible when both eigenvalues are real ($\omega=0$) \cite{ozturk2019fluctuation}, the decondensation threshold must always occur in the bi-exponential phase, c.f.~Fig.~\ref{fig:Fig_5}~(a),~(b). 

We furthermore notice that the strongly driven, non-equilibrium regime (upper right in Fig.~\ref{fig:Fig_3} (b)) represents a laser. That is, the near-equilibrium photon BEC phase is qualitatively separated from the laser phase by a hidden $g^{(2)}(t)$ transition, which can be circumvented via the path in parameter space shown in the inset of Fig.~\ref{fig:Fig_3} (b).

\begin{figure}[htb!]
    \centering
    \includegraphics[width=0.95\linewidth]{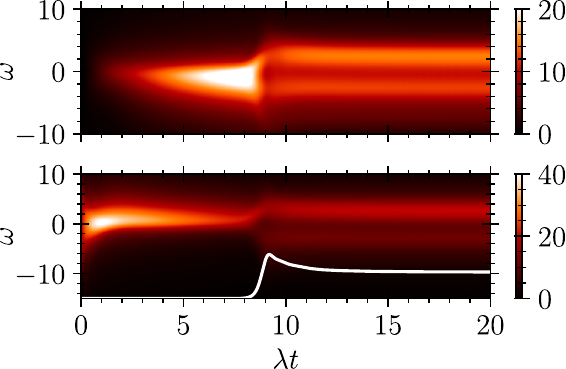}
    \caption{Photon self-energies $ -\mathrm{Im}\Sigma^{\gtrless}_{D}(t,\omega)$, i.e.\ effective emission (top) and absorption (bottom) coefficients, when pumping the system from the ground state. Parameters as in Fig.~\ref{fig:Fig_3} except for $g/\lambda = 2.45\cdot 10^{-4}$, $\Omega/\lambda = 1.25$, $\kappa/\lambda = 1.0 $ and $\bar{n} = 0.20$. Photon number shown in white (not to scale).}\label{fig:Fig_6}
\end{figure}

Finally, in Fig.~\ref{fig:Fig_6}, we study the non-equilibrium dynamics of the effective molecular emission and absorption spectra, as seen by the photons, for stronger coupling and at lower temperature. These results are obtained without further approximation such as memory truncation. Our method is thus capable of temporally resolving the emergence of characteristic spectral features (s.\ also Appendix~\ref{app:em_abs_spectra}).

\section{Conclusion.} 

We have introduced a non-equilibrium auxiliary field theory which faithfully describes the time-dependent quantum dynamics of general coupled system-bath set-ups, here generalized to open, driven-dissipative quantum systems such as photon BECs coupled to dye-molecule reservoirs~\cite{klaers2010bose} or exciton-polariton systems~\cite{Littlewood2006}. Our method may also be applied to the full quantum dynamics of multiple qubits coupled to sources of non-Markovian noise~\cite{White2020}.

For the open photon-BEC system, we find significant non-Markovian memory effects in the strongly driven regime where system loss $\kappa$ and reservoir relaxation $\lambda$ become comparable. We uncovered the global shape of the phase diagram partially explored in \cite{ozturk2020observation}. These calculations establish that the near-equilibrium photon BEC is separated from the lasing regime by a hidden phase transition of the photon density response.   

\begin{acknowledgments}
We acknowledge useful discussions with Fahri E. \"Ozt\"urk, Julian Schmitt, Michael Turaev, Frank Vewinger, and Martin Weitz. This work was supported in part by the Deutsche Forschungsgemeinschaft (DFG) within the Cooperative Research Center TRR 185 (277625399) and the Cluster of Excellence ML4Q (390534769).
\end{acknowledgments}

% \onecolumngrid

\appendix

\section{Rate Equations}\label{app:rate_eqs}

The rate equations for the photon and molecule number dynamics~\cite{Kirton2013, Kirton2015} can be derived from the full quantum field theory under the assumption that (in the frame rotating with the detuning $\delta_k$) the photon Green functions are approximately constant over the support of the molecule memory integrals, which is determined by the phonon relaxation rate $\lambda$. The two-time photon Green functions obey the equations of motion~ (see ~Eq.~(7) of the main text)
\myEq{
    \ii \partial_t  \boldsymbol{D}^\lessgtr(t, t') &= M\mint{t_0}{t}{\Bar{t}} \sbs{ \Sigma^>_D(t, \Bar{t}) -   \Sigma^<_D(t, \Bar{t})}   \boldsymbol{D}^\lessgtr(\Bar{t}, t') \\
    &- M\mint{t_0}{t'}{\Bar{t}}   \Sigma^\lessgtr_D(t, \Bar{t}) \sbs{ \boldsymbol{D}^>(\Bar{t}, t') -   \boldsymbol{D}^<(\Bar{t}, t')}
}
and 
\myEq{
        \ii \partial_{t'} \boldsymbol{D}^\lessgtr(t, t') &= M\mint{t_0}{t'}{\Bar{t}} \sbs{ \Sigma^>_D(\Bar{t}, t') - \Sigma^<_D(\Bar{t}, t')}   \boldsymbol{D}^\lessgtr(t, \Bar{t}) \\
        &- M\mint{t_0}{t}{\Bar{t}}   \Sigma^\lessgtr_D(\Bar{t}, t') \sbs{ \boldsymbol{D}^>(t, \Bar{t}) -   \boldsymbol{D}^<(t, \Bar{t})}.
}
Adding these two, one obtains a symmetrized equation of motion in terms of the center-of-motion time $T=(t+t')/2$ in the equal-time limit $t = t'$:
\myEq{\label{eq:aux_bosons_D_equal_time}
    \ii \partial_t  \boldsymbol{D}^<(t, t) = M\mint{t_0}{t}{\Bar{t}} \big[ &\Sigma^>_D(t, \Bar{t})   \boldsymbol{D}^<(\Bar{t}, t) \\[-0.2cm]
    - &\Sigma^<_D(t, \Bar{t})  \boldsymbol{D}^>(\Bar{t}, t) \\
    + &\boldsymbol{D}^<(t, \Bar{t}) \Sigma^>_D(\Bar{t}, t) \\
    - &\boldsymbol{D}^>(t, \Bar{t}) \Sigma^<_D(\Bar{t}, t)  \big].    
}
Then we can write the equation of motion for the photon number $N_{k}(t) = \ii D^<_{kk}(t,t)$, including the Lindblad relaxation term due to the cavity loss  $\kappa$ discussed in the main text, as 
\onecolumngrid
\myEq{\label{eq:aux_bosons_rate_eq}
\ii\partial_t D^<_{kk}(t, t) &=-\ii\kappa  D^<_{kk}(t, t) + \ii Mg^2 \mint{0}{t}{\Bar{t}}\Big\{\operatorname{Tr}\sbs{\boldsymbol{E}^>(t, \Bar{t})\boldsymbol{G}^<(\Bar{t}, t)}{D}^<_{kk}(\Bar{t}, t) + \operatorname{Tr}\sbs{\boldsymbol{G}^<(t, \Bar{t})\boldsymbol{E}^>(\Bar{t}, t)}{D}^<_{kk}(t, \Bar{t}) \\
     & \hspace{4.45cm}- \operatorname{Tr}\sbs{\boldsymbol{E}^<(t, \Bar{t})\boldsymbol{G}^>(\Bar{t}, t)}{D}^>_{kk}(\Bar{t}, t) - \operatorname{Tr}\sbs{\boldsymbol{G}^>(t, \Bar{t})\boldsymbol{E}^<(\Bar{t}, t)}{D}^>_{kk}(t, \Bar{t})\Big\},
}
\twocolumngrid\noindent 
where we have absorbed the prefactor $\zeta^{-1}$ of $\Sigma_D^{\lessgtr}$. Now, when the vibrational states have a rapid relaxation (i.e.\ are strongly broadened by collisions with the surrounding solvent), the dynamics of the auxiliary particles in relative time $(t-t')$ is approximately stationary and independent of the dynamics in forward time $T$. The vibrational states can then only have an overall forward evolution, which is to say that a quantum going either into the ground- or excited-state manifold will end up in vibrational state $n$ with a fixed probability $p_n$, where $\sum_n p_n = 1$. This enables us to write, for instance, 
\onecolumngrid
\myEq{\label{eq:auxiliary_bosons_g_nm}
   \mint{0}{t}{\Bar{t}} G^<_{nn'}(t, \bar{t}) &\approx G^<(t, t) \mint{0}{\infty}{(t - \bar{t})} \ee^{\ii\omega_D(t - \bar{t})} \ee^{-\rbs{\Gamma_\uparrow + \Gamma_\downarrow}(t-\bar{t})/2}g^{}_{nn'}(t - \bar{t}), \\
    \mint{0}{t}{\Bar{t}} E^<_{nn'}(t, \bar{t}) &\approx E^<(t, t) \mint{0}{\infty}{(t - \bar{t})} \ee^{\ii\omega_D(t - \bar{t})} \ee^{-\rbs{\Gamma_\uparrow + \Gamma_\downarrow}(t-\bar{t})/2}e^{}_{nn'}(t - \bar{t})   
}
\twocolumngrid\noindent
where $g^{}_{nn'}$ encodes the relative-time dynamics of $G^<_{nn'}(t, t')$ beyond the effect of $\omega_D$ and $\Gamma_{\uparrow,\,\downarrow}$, that is, the vibrational frequencies and relaxation. It has the property $g^{}_{nn'}(0) = \myFrac{G^<_{nn'}(t, t)}{G^<(t, t)}$, such that $g^{}_{nn}(0) = p^{}_n$. The analogous expression for the excited states is $e^{}_{nn'}(0) = \myFrac{E^<_{nn'}(t, t)}{E^<(t, t)}$. We furthermore introduce the total number of excited molecules as
\myEq{
    M_\uparrow(t) &= \ii M \operatorname{Tr}\boldsymbol{E}^<(t, t) = \ii M \sum_{n=0}^\infty E^<_{nn}(t, t) = \ii M E^<(t, t),
}
and similarly for the ground state where
\myEq{
    M - M_\uparrow(t) &= \ii M \operatorname{Tr}\boldsymbol{G}^<(t, t) \\
    &= \ii M \sum_{n=0}^\infty G^<_{nn}(t, t) = \ii M G^<(t, t),
}
such that the constraint $\hat{Q} = 1$ ensures total number conservation. Finally, the above assumptions ensure that the greater functions do not possess a forward-time dependence, i.e.\
\myEq{
    E^>_{nn'}(t, t') &\approx E^>_{nn'}(t - t'), \\
    G^>_{nn'}(t, t') &\approx G^>_{nn'}(t - t').
}
With these definitions, and using that under the approximation described above one can pull out the photon propagators from the time integrals and normalise by the respective occupation number of the ground and the excited states, we may transform Eq.~\eqref{eq:aux_bosons_rate_eq} into
\myEq{\label{eq:photons_rate_eq}
    \partial_t N_{k}(t) &= -\kappa N_k(t) \\
    &\phantom{=} -\sbs{K(+\delta_k) + K^*(+\delta_k)} N_{k}(t) \rbs{M - {M_\uparrow}(t)} \\
    &\phantom{=} + \sbs{K^*(-\delta_k) + K(-\delta_k)} \rbs{N_{k}(t) + 1} {M_\uparrow}(t) \\
    &= -\kappa N_k(t) -\Gamma^+_k\;N_k(t)\rbs{M - {M_\uparrow}(t)} \\
    &+ \Gamma^-_k \rbs{N_k(t) + 1}{M_\uparrow}(t),
}
which is the established result~\cite{Kirton2013, Kirton2015}, and we have identified $\Gamma^\pm_k = \Gamma(\pm\delta_k) = 2\operatorname{Re}K(\pm\delta_k)$. In our case, the emission and absorption coefficients are then given via
\myEq{\label{eq:supp_K_delta}
    K(\delta) = g^2 \mint{0}{\infty}{(t - \bar{t})} \ee^{\ii\delta(t - \bar{t})}\ee^{-\rbs{\Gamma_\uparrow + \Gamma_\downarrow}(t - \bar{t})/2}   A(t - \bar{t}),
}
where 
\myEq{
    A(\tau) &=\sum_{n,n'=0}^\infty\ii G^>_{nn'}(\tau) e^{}_{n'n}(-\tau) \\
    &= \sum_{n,n'=0}^\infty\ii E^>_{nn'}(\tau) g^{}_{n'n}(-\tau).
}
This is to be compared with the original expression in the quantum master equation derived via the usual Born-Markov approximation~\cite{Kirton2013, Kirton2015}:
\myEq{
    K(\delta) = g^2 \mint{0}{\infty}{(t - \bar{t})} \ee^{\ii\delta(t - \bar{t})}\ee^{-\rbs{\Gamma_\uparrow + \Gamma_\downarrow}(t - \bar{t})/2} f(t-\bar{t}),
}
where $f(t)$ is the polaron correlation function. An alternative and more practical way of defining Eq.~\eqref{eq:supp_K_delta}, which is also leading directly to Eq.~\eqref{eq:photons_rate_eq}, follows by letting
\myEq{\label{eq:supp_K_delta_II}
    K(\delta) &= \ii g^2 \rbs{\operatorname{Tr}\boldsymbol{G}^<(t, t)}^{-1} \mint{0}{t}{t} \operatorname{Tr}\sbs{\boldsymbol{{E}}^>(t, \Bar{t})\boldsymbol{{G}}^<(\Bar{t}, t)} \\
    &= \ii g^2 \rbs{\operatorname{Tr}\boldsymbol{E}^<(t, t)}^{-1} \mint{0}{t}{t} \operatorname{Tr}\sbs{\boldsymbol{{E}}^<(t, \Bar{t})\boldsymbol{{G}}^>(\Bar{t}, t)}.
}
Defining $K(\delta)$ in \textit{this} way occurs with the thought in mind that it is effectively the right-hand sides of Eq.\ \eqref{eq:supp_K_delta_II} acting instead of their previous definitions \cite{Kirton2013, Kirton2015} when parameters are chosen such that the auxiliary-particle dynamics effectively recovers the rate equations. Finally, the molecular rate equation follows completely analogously and reads
\begin{align}
    \begin{split}\label{eq:g2_m}
        \partial_t{ M_\uparrow(t) } &= \Gamma_\uparrow\left(M -  M_\uparrow(t)  \right) - \Gamma_\downarrow M_\uparrow(t) \\[0.05cm]
        &+ \sum_k\big[ \Gamma^+_k  N_k(t)  \left(M -  M_\uparrow(t)  \right) \\[-0.2cm]
        &\hspace{0.7cm}- \Gamma^-_k \left(N_k(t)  + 1 \right) M_\uparrow(t) \big].
    \end{split}
\end{align}

\section{Driven-Dissipative Processes in Schwinger-Keldysh Formalism}

To be self-contained, here we give a brief introduction to the Schwinger-Keldysh formalism as it applies to open systems. Subsequently, we show how to derive those parts of the equations of motions for the greater and lesser Green functions deriving from the introduction of driven-dissipative processes. We emphasize that we make no approximations here other than the formal diagrammatic expansion in the auxiliary particles. The projection as such is exact. Furthermore, the resulting equations of motion turn out to be effectively \textit{quadratic}, which hints at the fact that their validity is rather robust also for strong drive and dissipation.

As the simplest example, consider a cavity mode of frequency $\omega_0$, coupled to an environment at a low temperature such that $\hbar\omega_0\gg \beta^{-1}$, is described by the Lindblad master equation 
\begin{align}\label{eq:supp_lindblad_loss}
    \partial_{t} \hat{\rho}=-\ii\left[\hat{H} \hat{\rho}-\hat{\rho} \hat{H}^{
    \dagger}\right]+\kappa a \hat{\rho}  a^{\dagger},
\end{align}
with the non-Hermitian Hamiltonian
\begin{align}
    \hat{H}=\left(\omega_{0}-\ii \kappa/2\right) a^{\dagger} a.
\end{align}
In its most general formulation as given by Schwinger \cite{Schwinger1961}, the formalism easily captures this kind of dynamics. The Schwinger action that appears in the coherent-state expansion of the non-equilibrium partition function $Z$ is \cite{sieberer2016keldysh}
\myEq{
    S[\phi_{\pm}^{*} , \phi_{\pm}^{}] &= \int \text{d}{{t}} \left[ \phi_{+}^{*}\rbs{\ii\partial_{{t}} - \omega_0 + \ii\kappa/2} \phi_{+}^{} \right. \\
    &\left.- \phi_{-}^{*} \rbs{\ii \partial_{{t}} - \omega_0 - \ii\kappa/2} \phi_{-}^{}  -\ii \kappa \phi_{+}^{}\phi_{-}^{*} \right].
}
Note how this contains a contribution across the two branches of the contour. The respective equations of motion for the greater and lesser Green functions are then
\begin{align}\begin{split}
    0 &= \rbs{\ii \partial_t - \omega_0 + \ii\kappa/2 } D^<(t, t') ,   \\
    0 &= \rbs{\ii \partial_t - \omega_0 - \ii\kappa/2 } D^>(t, t')    + \ii\kappa D^T(t, t')  ,
\end{split}\end{align}
where now the time-ordered Green function appears explicitly. The anti-time-ordered Green function would appear, for instance, when coupling to a bath at finite temperature. In the equal-time limit, these equations become
\begin{align}\begin{split}\label{eq:supp_equal_time_greater_lesser}
    \partial_t D^<(t, t) &= -\kappa D^<(t, t) ,\\
    \partial_t D^>(t, t) &= \kappa D^>(t, t)  - \kappa\sbs{D^T(t, t) + D^{\tilde{T}}(t, t) }\\
    &= \kappa D^>(t, t)  - \kappa\sbs{D^>(t, t) + D^<(t, t) },
\end{split}\end{align}
which serves to illustrate how the commutator is preserved over time by the quantum jumps.

To have a consistent diagrammatic expansion for the four vertices $g$, $\Gamma_{\uparrow,\,\downarrow}$ and $\lambda$ of our theory, we expand the corresponding vertices to {second order in $\hbar$}. For the incoherent couplings, such a {two-loop expansion} amounts to working in Hartree-Fock approximation. 

\section{Determining the Oscillation Frequency and Decay Rates of the Second-Order Correlations}\label{app:fit}

\begin{figure*}
	\centering
	\includegraphics[width=0.8\linewidth]{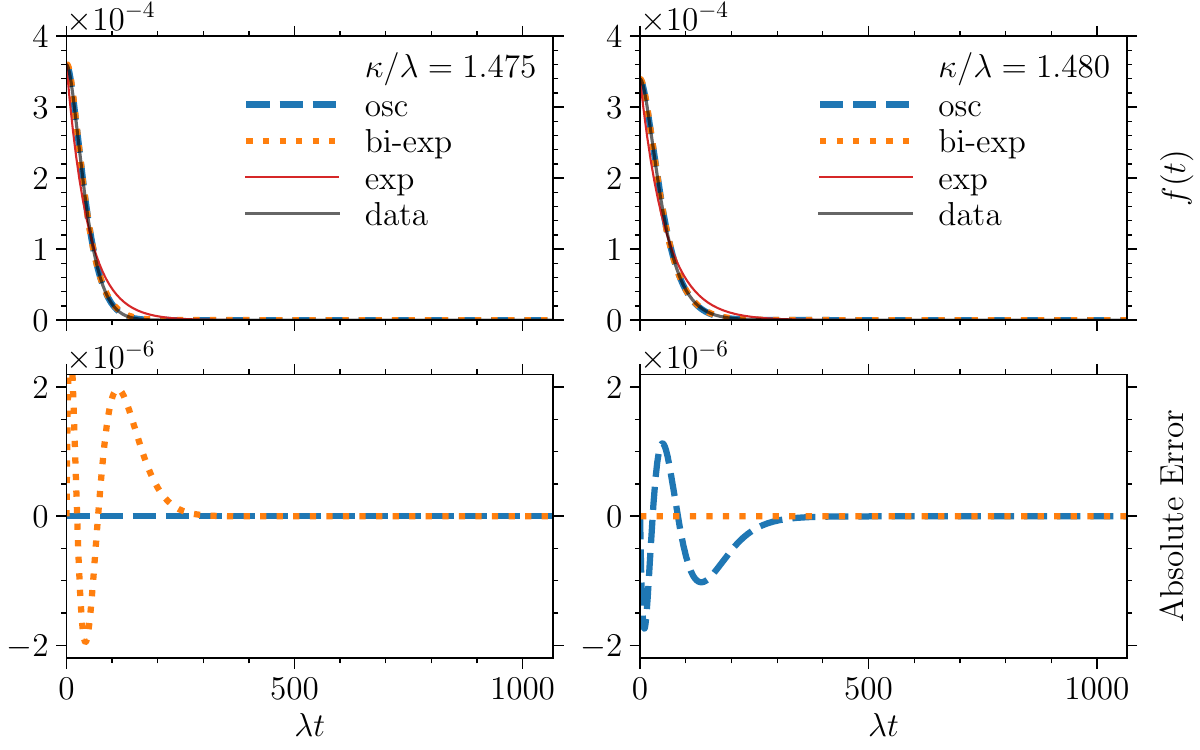}
	\caption{Comparison of the fits directly to the left and right of the transition at large $\kappa$, Upper panels: Data and best fits for several different ansatz functions. A uni-exponential decay is also fitted to underline that it is not a possible best fit, as can indeed be judged by eye. Lower panels: The absolute error (difference of data and fit) for the two viable options $f_{\text{osc}}$ and $f_{\text{bi-exp}}$.}
\label{fig:lsq_fit}
\end{figure*}

The density responses in the main panel of Fig.~3 (a) show the evolution of $\ii D^<(t, t)$ following the Gaussian pulse \eqref{eq:S_gauss}, where the point $t=0$ of the plot corresponds to the maximum of the curve after the Gaussian pulse has been injected. The frequency $\omega$ and decay rate $\tau$ in the oscillatory phase and the decay rates $\tau_1$, $\tau_2$ are extracted by least-squares fitting the functions
\myEq{\label{eq:supp_fit_funcs}
    f_{\text{osc}}(t) &= \ee^{-t/\tau}\rbs{f(0)\cos{\omega t} + C\sin{\omega t}}, \\
    f_{\text{bi-exp}}(t) &= \frac{f(0)}{2}{(\ee^{-t/\tau_1} + \ee^{-t/\tau_2})}   + \frac{C}{2}{(\ee^{-t/\tau_1} - \ee^{-t/\tau_2})}
}
to the numerical density responses. The quality of the fit is then estimated by the standard error, and the oscillatory or bi-exponential fit is accepted, whichever has the smaller error. This determines whether the response is classified as oscillating or bi-exponentially relaxing. An illustration of the fitting procedure is given in Fig.\ \ref{fig:lsq_fit} and Tabs. \ref{tab:lsq_fit_osc}, \ref{tab:lsq_fit_bi_exp}. One can see that even for these two points directly next to the transition, the fit classification is always unique.

\begin{table}
\centering
\begin{tabular}{c|c|c|c}
$\kappa/\lambda$ & $s(\tau)$            & $s(\omega)$           & $s(C)$                 \\ \hline
$1.475$  & $1.01\cdot 10^{-8}$ & $1.08\cdot 10^{-8}$  & $1.29\cdot 10^{-9}$ \\ \hline
$1.480$  & $2.35\cdot 10^{-5}$ & $1.18 \cdot 10^{-3}$ & $1.61\cdot 10^{-1}$ \\
\end{tabular}
\caption{Standard errors for the damped-oscillating ansatz.}
\label{tab:lsq_fit_osc}
\end{table}

\begin{table}
\centering
\begin{tabular}{c|c|c|c}
$\kappa/\lambda$ & $s(\tau_1)$          & $s(\tau_2)$           & $s(C)$                 \\ \hline
$1.475$          & $2.84\cdot 10^{-3}$ & $2.93 \cdot 10^{-3}$ & $2.85\cdot 10^{-1}$ \\ \hline
$1.480$          & $4.02\cdot 10^{-9}$ & $1.45 \cdot 10^{-8}$ & $4.42\cdot 10^{-10}$ \\ 
\end{tabular}
\caption{Standard errors for the bi-exponentially decaying ansatz.}
\label{tab:lsq_fit_bi_exp}
\end{table}

\section{Memory Truncation}\label{app:mem_trunc}

Solving the equations of motion for interacting systems numerically on the two-time grid becomes computationally expansive when the number of grid points needs to be large. This happens whenever the product of the fastest system frequency and the required final time is not small.
\begin{figure*}
	\centering
	\includegraphics[width=1.0\linewidth]{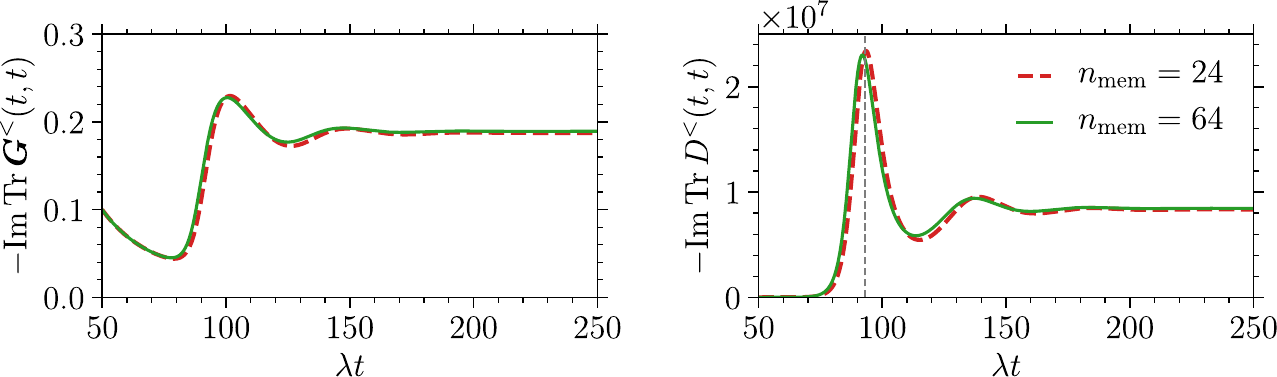}
	\caption{Forward dynamics of a single photon mode $k=0$ for different memory truncations. Initial conditions and parameters are $\operatorname{Tr} D^<(0,0)=0$, $\ii{G}^<_{00}(0, 0) = 1$, and $M=10^9$, $g/\lambda = 4.5\cdot 10^{-5}$, $\omega_0/\lambda = -1.0$, $\kappa/\lambda = 1.0$, $\Omega/\lambda = 1.0$, $S=1.0$, $\bar{n} = 0.25$, $\Gamma_\uparrow/\lambda = 0.05$, $\Gamma_\downarrow = \Gamma_\uparrow/40$, and vibrational-state truncation at $n=4$. Simulations computed for $2^{13}$ steps up to a final time $\lambda T_\text{max} = 512$ with memory times {$\lambda t_{\text{mem}} = \{1.5, 4.0\}$}. The steady-state occupations are $\lim_{t\to\infty}\ii\operatorname{Tr} D^<(t, t) = \{8.347\cdot 10^6, 8.446\cdot10^6\}$, which means that the short-memory value deviates only by about $0.012$.}
\label{fig:memory_truncation_T}
\end{figure*}
Any integral over relatively fast decaying Green functions, however, will be computable at sufficient precision over a small support that does not grow as the two-time grid expands towards it final size. In the same spirit, computing points far off the ``time diagonal'' $t = t'$ will be superfluous because they are negligible. In this way, by truncating the number of steps one moves away from the time diagonal, and by restricting the computation of the integrals to that same narrow ``band'', it is indeed possible to achieve a quasi-linear scaling in the number of grid points without losing accuracy. 

The quantitative consequences of different truncation parameters $n_{\text{mem}}$ are studied in Fig.\ \ref{fig:memory_truncation_T}. Note that technically, $n_{\text{mem}}$ is \textit{not} the number of points included away from the diagonal moving in the direction $\rbs{t - t'}$, but rather moving in the vertical and horizontal directions $t$ and $t'$. A short-valued memory $n_{\text{mem}} = 24$ results in a certain number of points inaccurately remaining zero. With a longer-valued memory $n_{\text{mem}}=64$, these points attain their proper values. However, looking to Fig.\ \ref{fig:memory_truncation_T}, we see that the influence of this is not considerable.

\section{Simulation of the Open Jaynes-Cummings Model}\label{app:open_JC}

\begin{figure}[htb!]
	\centering
	\includegraphics[width=0.95\linewidth]{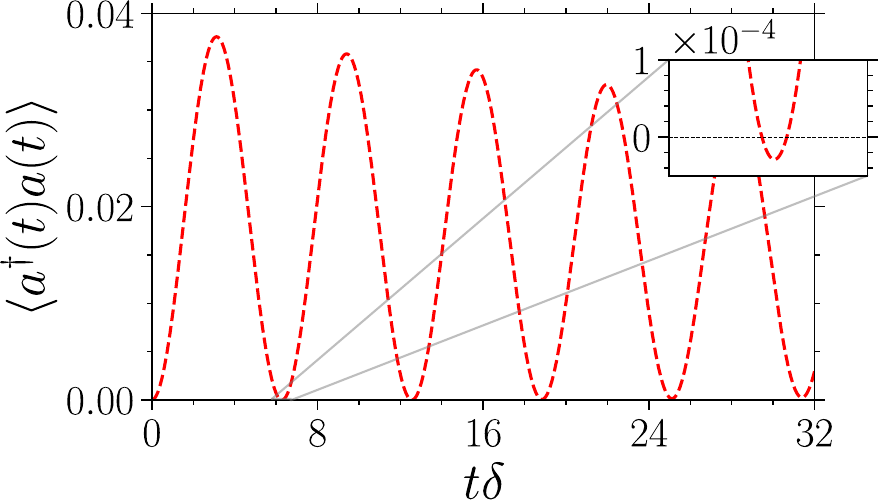}
	\caption{Solution of Eqs.\ \eqref{eq:open_JCM_trunc_eqs} with truncation $\E{a^\dagger a \sigma^z} \approx \E{a^\dagger a} \E{\sigma^z}$ for parameters $\kappa/\delta = 2^{-6}$, $g/\delta=2\pi \cdot 2^{-6}$. The negative occupation number is not a numerical artifact but results from the uncontrolled truncation.}
\label{fig:open_JCM_neg_num}
\end{figure}

A good playground for testing our methods is the simplest special case of the Hamiltonian in Eq.~(2) of the main paper, which is the Jaynes-Cummings model defined by
\myEq{\label{eq:slave_bosons_H_JC}
    H_{\text{JC}} &= \omega_0 a^{\dagger}_{} a^{}_{} + \frac{\Delta}{2} \sigma^z + g \left( a^{\dagger}_{} \sigma^- + a^{}_{}\sigma^+\right) \\
    &\equiv \delta a^{\dagger}_{} a^{}_{} + g \left( a^{\dagger}_{} \sigma^- + a^{}_{}\sigma^+\right),
}
where $\delta = \omega_0 - \Delta$. On top of the dynamics described by $H_{\text{JC}}$, we add a cavity loss for the photon mode as shown in the main text. From the standard master equation, one then derives the expectation-value equations of motion as
\myEq{\label{eq:open_JCM_trunc_eqs}
    \partial_t \E{a^\dagger a} &= -\kappa \E{a^\dagger a}  -\ii g \rbs{\E{a^\dagger \sigma^-} - \E{a \sigma^+}}, \\
    \partial_t \E{\sigma^z} &=  2\ii g \rbs{\E{a^\dagger \sigma^-} - \E{a \sigma^+}}, \\
    \partial_t \E{a\sigma^+} &= -\ii \delta \E{a\sigma^+} - \frac{\kappa}{2} \E{a\sigma^+} \\
    &- \ii g \rbs{\E{a^\dagger a \sigma^z} + \E{\sigma^+ \sigma^-}}.
}
These are, however, not closed because of the occurrence of $\E{a^\dagger a \sigma^z}$ in the last of Eqs.\ \eqref{eq:open_JCM_trunc_eqs}. In the literature, in generalizations of $H_{\text{JC}}$ to many spins, this term has been treated by means of the heuristic truncation $\E{a^\dagger a \sigma^z} \approx \E{a^\dagger a} \E{\sigma^z}$ \cite{Radonjic2018}, which is not a controlled approximation. As shown in Fig.\ \ref{fig:open_JCM_neg_num}, this can even result negative particle numbers (not a numerical artifact).
While this problem is mild for the parameters  chosen here, it becomes worse in other regimes. In particular, the factor of $M$ coming from the sum over many molecules in the equation for $\E{a^\dagger a}$ can lead to unphysical results. This is a manifestation of the difficulties involved in truncating the expectation-value hierarchy when phase coherence between photons and molecules is important. As explained above, the auxiliary-boson technique provides a resolution of this problem.
\begin{figure*}[htb!]
	\centering
	\includegraphics[width=1.0\linewidth]{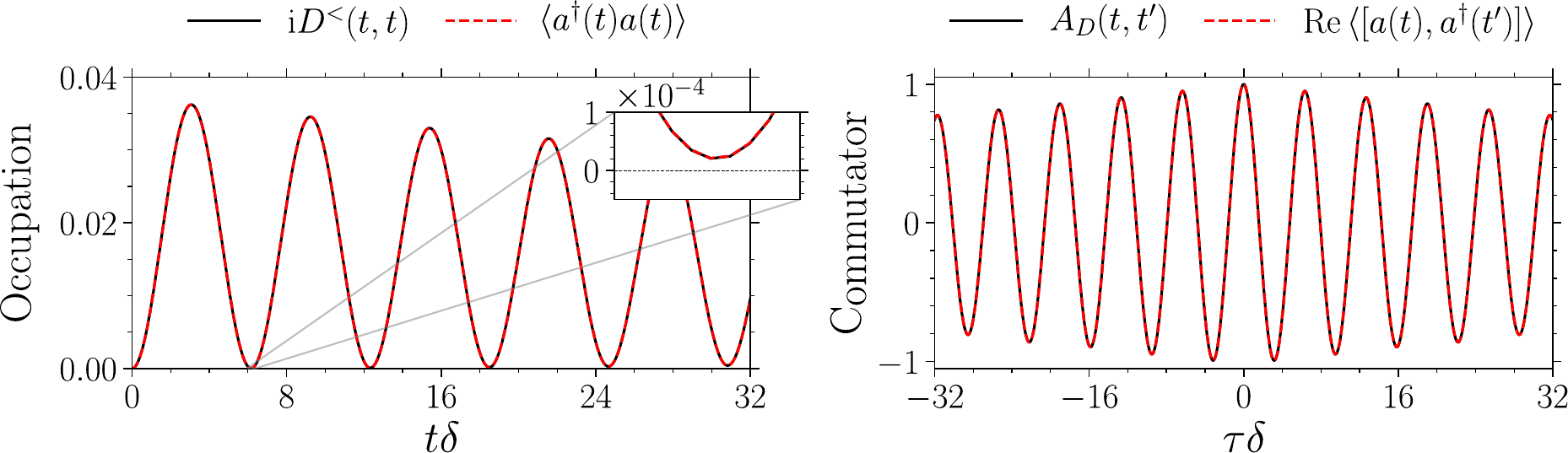}
	\caption{Solution of Eqs.~(6, 7) from the main text for parameters $\kappa/\delta = 2^{-6}$, $g/\delta=2\pi \cdot 2^{-6}$. The (red) dashed lines are the benchmark results calculated with a numerically exact method from the corresponding von Neumann equation. Because of the good agreement, the lines overlap identically. The spectral function is defined as $A_D(t, t') = -\operatorname{Im}\rbs{D^>(t, t') - D^<(t, t')}$. In the right panel, the times are $t = \rbs{T_\text{max} + \tau}/2$ and $t' = \rbs{T_\text{max} - \tau}/2$, where $T_\text{max}\delta =32.0$. The results were calculated with a step size of $\delta \mathrm{d}t = 10^{-2}$ for a number of $2^{10}$ steps.}
\label{fig:open_JCM_pos_num}
\end{figure*}
Here, this can be appreciated from Fig.\ \ref{fig:open_JCM_pos_num}. The inset highlights that the unphysical negative particle numbers have been removed by the auxiliary-particle method and that it reproduces numerically exact results. 

 \begin{figure}[htb!]
	\centering
	\includegraphics[width=0.95\linewidth]{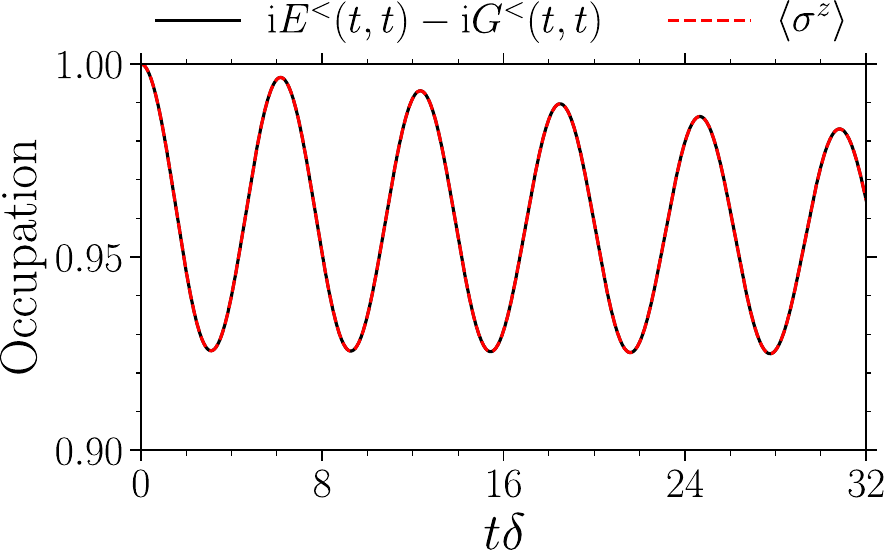}
	\caption{Equal-time occupation as given by the auxiliary-particle lesser Green functions and the corresponding numerically exact time evolution of $\E{\sigma^z}$. The parameters are those of Fig.\ \eqref{fig:open_JCM_pos_num}.}
\label{fig:open_JCM_sigma_z}
\end{figure}

\section{Time-Dependent Emission And Absorption Spectra}\label{app:em_abs_spectra}

\begin{figure*}[htb!]
    \centering
    \includegraphics[width=0.8\linewidth]{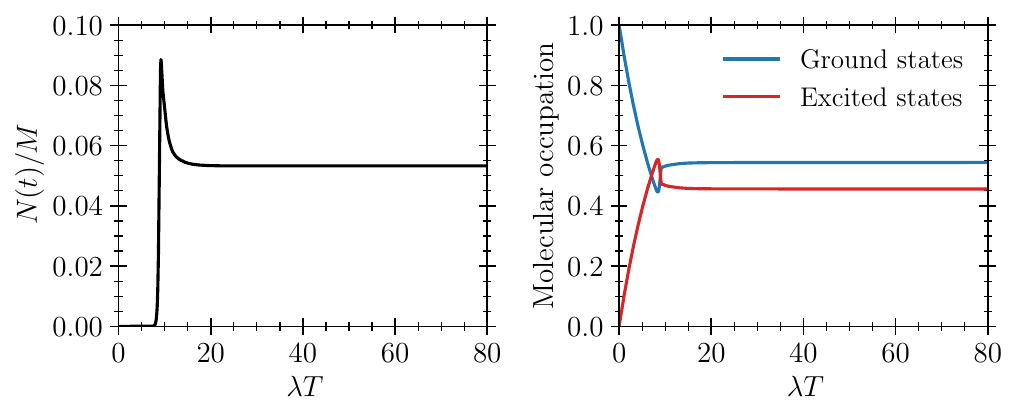}
    \caption{Photon occupation and molecular occupations as a function of time for initial conditions $N(0)=0$, $ \ii\operatorname{Tr} {G}^<(0, 0)  = 1$, and parameters  $M=10^9$, $g/\lambda = 24.5\cdot 10^{-5}$, $\delta_0/\lambda = -1.00$, $\Omega/\lambda = 1.25$, $S=1$, $\bar{n} = 0.2$, $\kappa/\lambda = 20$,$\Gamma_{\uparrow}/\lambda=1/20$ $\Gamma_\downarrow/\lambda = 1.25\cdot 10^{-3}$. We do not apply memory truncation and set numerical tolerances $\texttt{atol}=10^{-5}$ and $\texttt{rtol}=10^{-4}$ in the adaptive solver for Kadanoff-Baym equations~\cite{Meirinhos2022}.}\label{fig:Fig_12}
\end{figure*}

\begin{figure*}[htb!]
    \centering
    \includegraphics[width=0.8\linewidth]{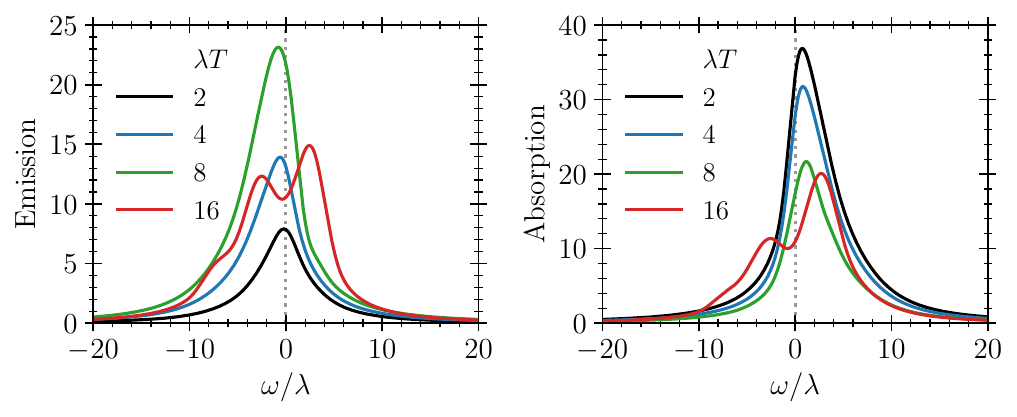}
    \caption{Photon self-energies $ -\mathrm{Im}\Sigma^{\gtrless}_{D}(T,\omega)$ (effective emission (left) and absorption (right) coefficients). The curves shown here are cross-sections of Fig.~4 in the main text.}\label{fig:Fig_13}
\end{figure*}

The photon self-energies as defined in the main text, i.e.\
\myEq{
    {\Sigma}_D^{\gtrless}(t, t') &= \ii g^{2}\zeta^{-1} \operatorname{Tr}{\boldsymbol{E}^{\gtrless}(t,t')\boldsymbol{G}^{\lessgtr}(t', t)},
}
can be used to define the effective emission and absorption spectra via which the molecules act on the photons. This can also be understood by comparing the above expression for the self-energies with Eq.~\eqref{eq:supp_K_delta_II}, which we defined as
\myEq{
    K(\delta) &= \ii g^2 \rbs{\operatorname{Tr}\boldsymbol{G}^<(t, t)}^{-1} \mint{0}{t}{t} \operatorname{Tr}\sbs{\boldsymbol{{E}}^>(t, \Bar{t})\boldsymbol{{G}}^<(\Bar{t}, t)}\\
    &= \ii g^2 \rbs{\operatorname{Tr}\boldsymbol{E}^<(t, t)}^{-1} \mint{0}{t}{t} \operatorname{Tr}\sbs{\boldsymbol{{E}}^<(t, \Bar{t})\boldsymbol{{G}}^>(\Bar{t}, t)}.
}
Here we show more data complementing Fig.~4 in the main text. The evolution toward the steady state of the occupation numbers if shown in Fig.~\ref{fig:Fig_12}, while additional cross-section through the effective spectra presented in the main text's Fig.~4 are given in Fig.~\ref{fig:Fig_13}. To obtain these curves, the self-energies are first evolved with their full dependence on $(t, t')$ and then transformed to so-called Wigner coordinates $T = (t+t')/2$, $\tau = t - t'$. The latter relative-time variable is then Fourier transformed to obtain the spectra shown in Fig.~\ref{fig:Fig_13}.

\bibliography{refs}

\begin{thebibliography}{50}%
\makeatletter
\providecommand \@ifxundefined [1]{%
 \@ifx{#1\undefined}
}%
\providecommand \@ifnum [1]{%
 \ifnum #1\expandafter \@firstoftwo
 \else \expandafter \@secondoftwo
 \fi
}%
\providecommand \@ifx [1]{%
 \ifx #1\expandafter \@firstoftwo
 \else \expandafter \@secondoftwo
 \fi
}%
\providecommand \natexlab [1]{#1}%
\providecommand \enquote  [1]{``#1''}%
\providecommand \bibnamefont  [1]{#1}%
\providecommand \bibfnamefont [1]{#1}%
\providecommand \citenamefont [1]{#1}%
\providecommand \href@noop [0]{\@secondoftwo}%
\providecommand \href [0]{\begingroup \@sanitize@url \@href}%
\providecommand \@href[1]{\@@startlink{#1}\@@href}%
\providecommand \@@href[1]{\endgroup#1\@@endlink}%
\providecommand \@sanitize@url [0]{\catcode `\\12\catcode `\$12\catcode
  `\&12\catcode `\#12\catcode `\^12\catcode `\_12\catcode `\%12\relax}%
\providecommand \@@startlink[1]{}%
\providecommand \@@endlink[0]{}%
\providecommand \url  [0]{\begingroup\@sanitize@url \@url }%
\providecommand \@url [1]{\endgroup\@href {#1}{\urlprefix }}%
\providecommand \urlprefix  [0]{URL }%
\providecommand \Eprint [0]{\href }%
\providecommand \doibase [0]{http://dx.doi.org/}%
\providecommand \selectlanguage [0]{\@gobble}%
\providecommand \bibinfo  [0]{\@secondoftwo}%
\providecommand \bibfield  [0]{\@secondoftwo}%
\providecommand \translation [1]{[#1]}%
\providecommand \BibitemOpen [0]{}%
\providecommand \bibitemStop [0]{}%
\providecommand \bibitemNoStop [0]{.\EOS\space}%
\providecommand \EOS [0]{\spacefactor3000\relax}%
\providecommand \BibitemShut  [1]{\csname bibitem#1\endcsname}%
\let\auto@bib@innerbib\@empty
%</preamble>
\bibitem [{\citenamefont {Klaers}\ \emph {et~al.}(2010)\citenamefont {Klaers},
  \citenamefont {Schmitt}, \citenamefont {Vewinger},\ and\ \citenamefont
  {Weitz}}]{klaers2010bose}%
  \BibitemOpen
  \bibfield  {author} {\bibinfo {author} {\bibfnamefont {Jan}\ \bibnamefont
  {Klaers}}, \bibinfo {author} {\bibfnamefont {Julian}\ \bibnamefont
  {Schmitt}}, \bibinfo {author} {\bibfnamefont {Frank}\ \bibnamefont
  {Vewinger}}, \ and\ \bibinfo {author} {\bibfnamefont {Martin}\ \bibnamefont
  {Weitz}},\ }\bibfield  {title} {\enquote {\bibinfo {title} {Bose--einstein
  condensation of photons in an optical microcavity},}\ }\href {\doibase
  10.1038/nature09567} {\bibfield  {journal} {\bibinfo  {journal} {Nature}\
  }\textbf {\bibinfo {volume} {468}},\ \bibinfo {pages} {545--548} (\bibinfo
  {year} {2010})}\BibitemShut {NoStop}%
\bibitem [{\citenamefont {Klaers}\ \emph {et~al.}(2012)\citenamefont {Klaers},
  \citenamefont {Schmitt}, \citenamefont {Damm}, \citenamefont {Vewinger},\
  and\ \citenamefont {Weitz}}]{klaers2012statistical}%
  \BibitemOpen
  \bibfield  {author} {\bibinfo {author} {\bibfnamefont {Jan}\ \bibnamefont
  {Klaers}}, \bibinfo {author} {\bibfnamefont {Julian}\ \bibnamefont
  {Schmitt}}, \bibinfo {author} {\bibfnamefont {Tobias}\ \bibnamefont {Damm}},
  \bibinfo {author} {\bibfnamefont {Frank}\ \bibnamefont {Vewinger}}, \ and\
  \bibinfo {author} {\bibfnamefont {Martin}\ \bibnamefont {Weitz}},\ }\bibfield
   {title} {\enquote {\bibinfo {title} {{Statistical Physics of
  Bose-Einstein-Condensed Light in a Dye Microcavity}},}\ }\href {\doibase
  10.1103/PhysRevLett.108.160403} {\bibfield  {journal} {\bibinfo  {journal}
  {Phys. Rev. Lett.}\ }\textbf {\bibinfo {volume} {108}},\ \bibinfo {pages}
  {160403} (\bibinfo {year} {2012})}\BibitemShut {NoStop}%
\bibitem [{\citenamefont {Schmitt}\ \emph {et~al.}(2014)\citenamefont
  {Schmitt}, \citenamefont {Damm}, \citenamefont {Dung}, \citenamefont
  {Vewinger}, \citenamefont {Klaers},\ and\ \citenamefont
  {Weitz}}]{schmitt2014observation}%
  \BibitemOpen
  \bibfield  {author} {\bibinfo {author} {\bibfnamefont {Julian}\ \bibnamefont
  {Schmitt}}, \bibinfo {author} {\bibfnamefont {Tobias}\ \bibnamefont {Damm}},
  \bibinfo {author} {\bibfnamefont {David}\ \bibnamefont {Dung}}, \bibinfo
  {author} {\bibfnamefont {Frank}\ \bibnamefont {Vewinger}}, \bibinfo {author}
  {\bibfnamefont {Jan}\ \bibnamefont {Klaers}}, \ and\ \bibinfo {author}
  {\bibfnamefont {Martin}\ \bibnamefont {Weitz}},\ }\bibfield  {title}
  {\enquote {\bibinfo {title} {{Observation of Grand-Canonical Number
  Statistics in a Photon Bose-Einstein Condensate}},}\ }\href {\doibase
  10.1103/PhysRevLett.112.030401} {\bibfield  {journal} {\bibinfo  {journal}
  {Phys. Rev. Lett.}\ }\textbf {\bibinfo {volume} {112}},\ \bibinfo {pages}
  {030401} (\bibinfo {year} {2014})}\BibitemShut {NoStop}%
\bibitem [{\citenamefont {Hesten}\ \emph {et~al.}(2018)\citenamefont {Hesten},
  \citenamefont {Nyman},\ and\ \citenamefont
  {Mintert}}]{hesten2018decondensation}%
  \BibitemOpen
  \bibfield  {author} {\bibinfo {author} {\bibfnamefont {Henry~J.}\
  \bibnamefont {Hesten}}, \bibinfo {author} {\bibfnamefont {Robert~A.}\
  \bibnamefont {Nyman}}, \ and\ \bibinfo {author} {\bibfnamefont {Florian}\
  \bibnamefont {Mintert}},\ }\bibfield  {title} {\enquote {\bibinfo {title}
  {{Decondensation in Nonequilibrium Photonic Condensates: When Less Is
  More}},}\ }\href {\doibase 10.1103/PhysRevLett.120.040601} {\bibfield
  {journal} {\bibinfo  {journal} {Phys. Rev. Lett.}\ }\textbf {\bibinfo
  {volume} {120}},\ \bibinfo {pages} {040601} (\bibinfo {year}
  {2018})}\BibitemShut {NoStop}%
\bibitem [{\citenamefont {Walker}\ \emph {et~al.}(2018)\citenamefont {Walker},
  \citenamefont {Flatten}, \citenamefont {Hesten}, \citenamefont {Mintert},
  \citenamefont {Hunger}, \citenamefont {Trichet}, \citenamefont {Smith},\ and\
  \citenamefont {Nyman}}]{walker2018driven}%
  \BibitemOpen
  \bibfield  {author} {\bibinfo {author} {\bibfnamefont {Benjamin~T.}\
  \bibnamefont {Walker}}, \bibinfo {author} {\bibfnamefont {Lucas~C.}\
  \bibnamefont {Flatten}}, \bibinfo {author} {\bibfnamefont {Henry~J.}\
  \bibnamefont {Hesten}}, \bibinfo {author} {\bibfnamefont {Florian}\
  \bibnamefont {Mintert}}, \bibinfo {author} {\bibfnamefont {David}\
  \bibnamefont {Hunger}}, \bibinfo {author} {\bibfnamefont {Aur{\'e}lien
  A.~P.}\ \bibnamefont {Trichet}}, \bibinfo {author} {\bibfnamefont {Jason~M.}\
  \bibnamefont {Smith}}, \ and\ \bibinfo {author} {\bibfnamefont {Robert~A.}\
  \bibnamefont {Nyman}},\ }\bibfield  {title} {\enquote {\bibinfo {title}
  {Driven-dissipative non-equilibrium bose--einstein condensation of less than
  ten photons},}\ }\href {\doibase 10.1038/s41567-018-0270-1} {\bibfield
  {journal} {\bibinfo  {journal} {Nature Physics}\ }\textbf {\bibinfo {volume}
  {14}},\ \bibinfo {pages} {1173--1177} (\bibinfo {year} {2018})}\BibitemShut
  {NoStop}%
\bibitem [{\citenamefont {Keeling}\ and\ \citenamefont
  {K\'{e}na-Cohen}(2020)}]{keeling2020bose}%
  \BibitemOpen
  \bibfield  {author} {\bibinfo {author} {\bibfnamefont {Jonathan}\
  \bibnamefont {Keeling}}\ and\ \bibinfo {author} {\bibfnamefont
  {St\'{e}phane}\ \bibnamefont {K\'{e}na-Cohen}},\ }\bibfield  {title}
  {\enquote {\bibinfo {title} {{Bose–Einstein Condensation of
  Exciton-Polaritons in Organic Microcavities}},}\ }\href {\doibase
  10.1146/annurev-physchem-010920-102509} {\bibfield  {journal} {\bibinfo
  {journal} {Annual Review of Physical Chemistry}\ }\textbf {\bibinfo {volume}
  {71}},\ \bibinfo {pages} {435--459} (\bibinfo {year} {2020})},\ \bibinfo
  {note} {pMID: 32126177}\BibitemShut {NoStop}%
\bibitem [{\citenamefont {Ramezani}\ \emph {et~al.}(2019)\citenamefont
  {Ramezani}, \citenamefont {Halpin}, \citenamefont {Wang}, \citenamefont
  {Berghuis},\ and\ \citenamefont {Rivas}}]{ramezani2019ultrafast}%
  \BibitemOpen
  \bibfield  {author} {\bibinfo {author} {\bibfnamefont {Mohammad}\
  \bibnamefont {Ramezani}}, \bibinfo {author} {\bibfnamefont {Alexei}\
  \bibnamefont {Halpin}}, \bibinfo {author} {\bibfnamefont {Shaojun}\
  \bibnamefont {Wang}}, \bibinfo {author} {\bibfnamefont {Matthijs}\
  \bibnamefont {Berghuis}}, \ and\ \bibinfo {author} {\bibfnamefont
  {Jaime~G{\'o}mez}\ \bibnamefont {Rivas}},\ }\bibfield  {title} {\enquote
  {\bibinfo {title} {{Ultrafast Dynamics of Nonequilibrium Organic
  Exciton--Polariton Condensates}},}\ }\href {\doibase
  10.1021/acs.nanolett.9b03139} {\bibfield  {journal} {\bibinfo  {journal}
  {Nano Letters}\ }\textbf {\bibinfo {volume} {19}},\ \bibinfo {pages}
  {8590--8596} (\bibinfo {year} {2019})}\BibitemShut {NoStop}%
\bibitem [{\citenamefont {Hakala}\ \emph {et~al.}(2018)\citenamefont {Hakala},
  \citenamefont {Moilanen}, \citenamefont {V{\"a}kev{\"a}inen}, \citenamefont
  {Guo}, \citenamefont {Martikainen}, \citenamefont {Daskalakis}, \citenamefont
  {Rekola}, \citenamefont {Julku},\ and\ \citenamefont
  {T{\"o}rm{\"a}}}]{hakala2018bose}%
  \BibitemOpen
  \bibfield  {author} {\bibinfo {author} {\bibfnamefont {Tommi~K.}\
  \bibnamefont {Hakala}}, \bibinfo {author} {\bibfnamefont {Antti~J.}\
  \bibnamefont {Moilanen}}, \bibinfo {author} {\bibfnamefont {Aaro~I.}\
  \bibnamefont {V{\"a}kev{\"a}inen}}, \bibinfo {author} {\bibfnamefont {Rui}\
  \bibnamefont {Guo}}, \bibinfo {author} {\bibfnamefont {Jani-Petri}\
  \bibnamefont {Martikainen}}, \bibinfo {author} {\bibfnamefont
  {Konstantinos~S.}\ \bibnamefont {Daskalakis}}, \bibinfo {author}
  {\bibfnamefont {Heikki~T.}\ \bibnamefont {Rekola}}, \bibinfo {author}
  {\bibfnamefont {Aleksi}\ \bibnamefont {Julku}}, \ and\ \bibinfo {author}
  {\bibfnamefont {P{\"a}ivi}\ \bibnamefont {T{\"o}rm{\"a}}},\ }\bibfield
  {title} {\enquote {\bibinfo {title} {Bose--einstein condensation in a
  plasmonic lattice},}\ }\href {\doibase 10.1038/s41567-018-0109-9} {\bibfield
  {journal} {\bibinfo  {journal} {Nature Physics}\ }\textbf {\bibinfo {volume}
  {14}},\ \bibinfo {pages} {739--744} (\bibinfo {year} {2018})}\BibitemShut
  {NoStop}%
\bibitem [{\citenamefont {Clear}\ \emph {et~al.}(2020)\citenamefont {Clear},
  \citenamefont {Schofield}, \citenamefont {Major}, \citenamefont {Iles-Smith},
  \citenamefont {Clark},\ and\ \citenamefont {McCutcheon}}]{Clear2020}%
  \BibitemOpen
  \bibfield  {author} {\bibinfo {author} {\bibfnamefont {Chloe}\ \bibnamefont
  {Clear}}, \bibinfo {author} {\bibfnamefont {Ross~C.}\ \bibnamefont
  {Schofield}}, \bibinfo {author} {\bibfnamefont {Kyle~D.}\ \bibnamefont
  {Major}}, \bibinfo {author} {\bibfnamefont {Jake}\ \bibnamefont
  {Iles-Smith}}, \bibinfo {author} {\bibfnamefont {Alex~S.}\ \bibnamefont
  {Clark}}, \ and\ \bibinfo {author} {\bibfnamefont {Dara P.~S.}\ \bibnamefont
  {McCutcheon}},\ }\bibfield  {title} {\enquote {\bibinfo {title}
  {{Phonon-Induced Optical Dephasing in Single Organic Molecules}},}\ }\href
  {\doibase 10.1103/PhysRevLett.124.153602} {\bibfield  {journal} {\bibinfo
  {journal} {Phys. Rev. Lett.}\ }\textbf {\bibinfo {volume} {124}},\ \bibinfo
  {pages} {153602} (\bibinfo {year} {2020})}\BibitemShut {NoStop}%
\bibitem [{\citenamefont {Sieberer}\ \emph {et~al.}(2016)\citenamefont
  {Sieberer}, \citenamefont {Buchhold},\ and\ \citenamefont
  {Diehl}}]{sieberer2016keldysh}%
  \BibitemOpen
  \bibfield  {author} {\bibinfo {author} {\bibfnamefont {L~M}\ \bibnamefont
  {Sieberer}}, \bibinfo {author} {\bibfnamefont {M}~\bibnamefont {Buchhold}}, \
  and\ \bibinfo {author} {\bibfnamefont {S}~\bibnamefont {Diehl}},\ }\bibfield
  {title} {\enquote {\bibinfo {title} {Keldysh field theory for driven open
  quantum systems},}\ }\href {\doibase 10.1088/0034-4885/79/9/096001}
  {\bibfield  {journal} {\bibinfo  {journal} {Reports on Progress in Physics}\
  }\textbf {\bibinfo {volume} {79}},\ \bibinfo {pages} {096001} (\bibinfo
  {year} {2016})}\BibitemShut {NoStop}%
\bibitem [{\citenamefont {Öztürk}\ \emph {et~al.}(2021)\citenamefont
  {Öztürk}, \citenamefont {Lappe}, \citenamefont {Hellmann}, \citenamefont
  {Schmitt}, \citenamefont {Klaers}, \citenamefont {Vewinger}, \citenamefont
  {Kroha},\ and\ \citenamefont {Weitz}}]{ozturk2020observation}%
  \BibitemOpen
  \bibfield  {author} {\bibinfo {author} {\bibfnamefont {Fahri~Emre}\
  \bibnamefont {Öztürk}}, \bibinfo {author} {\bibfnamefont {Tim}\
  \bibnamefont {Lappe}}, \bibinfo {author} {\bibfnamefont {Göran}\
  \bibnamefont {Hellmann}}, \bibinfo {author} {\bibfnamefont {Julian}\
  \bibnamefont {Schmitt}}, \bibinfo {author} {\bibfnamefont {Jan}\ \bibnamefont
  {Klaers}}, \bibinfo {author} {\bibfnamefont {Frank}\ \bibnamefont
  {Vewinger}}, \bibinfo {author} {\bibfnamefont {Johann}\ \bibnamefont
  {Kroha}}, \ and\ \bibinfo {author} {\bibfnamefont {Martin}\ \bibnamefont
  {Weitz}},\ }\bibfield  {title} {\enquote {\bibinfo {title} {Observation of a
  non-hermitian phase transition in an optical quantum gas},}\ }\href {\doibase
  10.1126/science.abe9869} {\bibfield  {journal} {\bibinfo  {journal}
  {Science}\ }\textbf {\bibinfo {volume} {372}},\ \bibinfo {pages} {88}
  (\bibinfo {year} {2021})}\BibitemShut {NoStop}%
\bibitem [{\citenamefont {Ozturk}\ \emph {et~al.}(2019)\citenamefont {Ozturk},
  \citenamefont {Lappe}, \citenamefont {Hellmann}, \citenamefont {Schmitt},
  \citenamefont {Klaers}, \citenamefont {Vewinger}, \citenamefont {Kroha},\
  and\ \citenamefont {Weitz}}]{ozturk2019fluctuation}%
  \BibitemOpen
  \bibfield  {author} {\bibinfo {author} {\bibfnamefont {Fahri~Emre}\
  \bibnamefont {Ozturk}}, \bibinfo {author} {\bibfnamefont {Tim}\ \bibnamefont
  {Lappe}}, \bibinfo {author} {\bibfnamefont {G\"oran}\ \bibnamefont
  {Hellmann}}, \bibinfo {author} {\bibfnamefont {Julian}\ \bibnamefont
  {Schmitt}}, \bibinfo {author} {\bibfnamefont {Jan}\ \bibnamefont {Klaers}},
  \bibinfo {author} {\bibfnamefont {Frank}\ \bibnamefont {Vewinger}}, \bibinfo
  {author} {\bibfnamefont {Johann}\ \bibnamefont {Kroha}}, \ and\ \bibinfo
  {author} {\bibfnamefont {Martin}\ \bibnamefont {Weitz}},\ }\bibfield  {title}
  {\enquote {\bibinfo {title} {Fluctuation dynamics of an open photon
  bose-einstein condensate},}\ }\href {\doibase 10.1103/PhysRevA.100.043803}
  {\bibfield  {journal} {\bibinfo  {journal} {Phys. Rev. A}\ }\textbf {\bibinfo
  {volume} {100}},\ \bibinfo {pages} {043803} (\bibinfo {year}
  {2019})}\BibitemShut {NoStop}%
\bibitem [{\citenamefont {Schmitt}(2018)}]{Schmitt2018}%
  \BibitemOpen
  \bibfield  {author} {\bibinfo {author} {\bibfnamefont {Julian}\ \bibnamefont
  {Schmitt}},\ }\bibfield  {title} {\enquote {\bibinfo {title} {Dynamics and
  correlations of a bose–einstein condensate of photons},}\ }\href {\doibase
  10.1088/1361-6455/aad409} {\bibfield  {journal} {\bibinfo  {journal} {Journal
  of Physics B: Atomic, Molecular and Optical Physics}\ }\textbf {\bibinfo
  {volume} {51}},\ \bibinfo {pages} {173001} (\bibinfo {year}
  {2018})}\BibitemShut {NoStop}%
\bibitem [{\citenamefont {Kurtscheid}\ \emph {et~al.}(2019)\citenamefont
  {Kurtscheid}, \citenamefont {Dung}, \citenamefont {Busley}, \citenamefont
  {Vewinger}, \citenamefont {Rosch},\ and\ \citenamefont
  {Weitz}}]{Kurtscheid2019}%
  \BibitemOpen
  \bibfield  {author} {\bibinfo {author} {\bibfnamefont {Christian}\
  \bibnamefont {Kurtscheid}}, \bibinfo {author} {\bibfnamefont {David}\
  \bibnamefont {Dung}}, \bibinfo {author} {\bibfnamefont {Erik}\ \bibnamefont
  {Busley}}, \bibinfo {author} {\bibfnamefont {Frank}\ \bibnamefont
  {Vewinger}}, \bibinfo {author} {\bibfnamefont {Achim}\ \bibnamefont {Rosch}},
  \ and\ \bibinfo {author} {\bibfnamefont {Martin}\ \bibnamefont {Weitz}},\
  }\bibfield  {title} {\enquote {\bibinfo {title} {Thermally condensing photons
  into a coherently split state of light},}\ }\href {\doibase
  10.1126/science.aay1334} {\bibfield  {journal} {\bibinfo  {journal}
  {Science}\ }\textbf {\bibinfo {volume} {366}},\ \bibinfo {pages} {894--897}
  (\bibinfo {year} {2019})}\BibitemShut {NoStop}%
\bibitem [{\citenamefont {de~Vega}\ and\ \citenamefont
  {Alonso}(2017)}]{de2017dynamics}%
  \BibitemOpen
  \bibfield  {author} {\bibinfo {author} {\bibfnamefont {In\'es}\ \bibnamefont
  {de~Vega}}\ and\ \bibinfo {author} {\bibfnamefont {Daniel}\ \bibnamefont
  {Alonso}},\ }\bibfield  {title} {\enquote {\bibinfo {title} {Dynamics of
  non-markovian open quantum systems},}\ }\href {\doibase
  10.1103/RevModPhys.89.015001} {\bibfield  {journal} {\bibinfo  {journal}
  {Rev. Mod. Phys.}\ }\textbf {\bibinfo {volume} {89}},\ \bibinfo {pages}
  {015001} (\bibinfo {year} {2017})}\BibitemShut {NoStop}%
\bibitem [{\citenamefont {del Pino}\ \emph {et~al.}(2018)\citenamefont {del
  Pino}, \citenamefont {Schr\"oder}, \citenamefont {Chin}, \citenamefont
  {Feist},\ and\ \citenamefont {Garcia-Vidal}}]{del2018tensor}%
  \BibitemOpen
  \bibfield  {author} {\bibinfo {author} {\bibfnamefont {Javier}\ \bibnamefont
  {del Pino}}, \bibinfo {author} {\bibfnamefont {Florian A. Y.~N.}\
  \bibnamefont {Schr\"oder}}, \bibinfo {author} {\bibfnamefont {Alex~W.}\
  \bibnamefont {Chin}}, \bibinfo {author} {\bibfnamefont {Johannes}\
  \bibnamefont {Feist}}, \ and\ \bibinfo {author} {\bibfnamefont
  {Francisco~J.}\ \bibnamefont {Garcia-Vidal}},\ }\bibfield  {title} {\enquote
  {\bibinfo {title} {{Tensor Network Simulation of Non-Markovian Dynamics in
  Organic Polaritons}},}\ }\href {\doibase 10.1103/PhysRevLett.121.227401}
  {\bibfield  {journal} {\bibinfo  {journal} {Phys. Rev. Lett.}\ }\textbf
  {\bibinfo {volume} {121}},\ \bibinfo {pages} {227401} (\bibinfo {year}
  {2018})}\BibitemShut {NoStop}%
\bibitem [{\citenamefont {Kirton}\ and\ \citenamefont
  {Keeling}(2013)}]{Kirton2013}%
  \BibitemOpen
  \bibfield  {author} {\bibinfo {author} {\bibfnamefont {Peter}\ \bibnamefont
  {Kirton}}\ and\ \bibinfo {author} {\bibfnamefont {Jonathan}\ \bibnamefont
  {Keeling}},\ }\bibfield  {title} {\enquote {\bibinfo {title} {{Nonequilibrium
  Model of Photon Condensation}},}\ }\href {\doibase
  10.1103/PhysRevLett.111.100404} {\bibfield  {journal} {\bibinfo  {journal}
  {Phys. Rev. Lett.}\ }\textbf {\bibinfo {volume} {111}},\ \bibinfo {pages}
  {100404} (\bibinfo {year} {2013})}\BibitemShut {NoStop}%
\bibitem [{\citenamefont {Kirton}\ and\ \citenamefont
  {Keeling}(2015)}]{Kirton2015}%
  \BibitemOpen
  \bibfield  {author} {\bibinfo {author} {\bibfnamefont {Peter}\ \bibnamefont
  {Kirton}}\ and\ \bibinfo {author} {\bibfnamefont {Jonathan}\ \bibnamefont
  {Keeling}},\ }\bibfield  {title} {\enquote {\bibinfo {title} {Thermalization
  and breakdown of thermalization in photon condensates},}\ }\href {\doibase
  10.1103/PhysRevA.91.033826} {\bibfield  {journal} {\bibinfo  {journal} {Phys.
  Rev. A}\ }\textbf {\bibinfo {volume} {91}},\ \bibinfo {pages} {033826}
  (\bibinfo {year} {2015})}\BibitemShut {NoStop}%
\bibitem [{\citenamefont {Keeling}\ and\ \citenamefont
  {Kirton}(2016)}]{Keeling2016}%
  \BibitemOpen
  \bibfield  {author} {\bibinfo {author} {\bibfnamefont {Jonathan}\
  \bibnamefont {Keeling}}\ and\ \bibinfo {author} {\bibfnamefont {Peter}\
  \bibnamefont {Kirton}},\ }\bibfield  {title} {\enquote {\bibinfo {title}
  {Spatial dynamics, thermalization, and gain clamping in a photon
  condensate},}\ }\href {\doibase 10.1103/PhysRevA.93.013829} {\bibfield
  {journal} {\bibinfo  {journal} {Phys. Rev. A}\ }\textbf {\bibinfo {volume}
  {93}},\ \bibinfo {pages} {013829} (\bibinfo {year} {2016})}\BibitemShut
  {NoStop}%
\bibitem [{\citenamefont {Radonji{\'{c}}}\ \emph {et~al.}(2018)\citenamefont
  {Radonji{\'{c}}}, \citenamefont {Kopylov}, \citenamefont {Bala{\v{z}}},\ and\
  \citenamefont {Pelster}}]{Radonjic2018}%
  \BibitemOpen
  \bibfield  {author} {\bibinfo {author} {\bibfnamefont {Milan}\ \bibnamefont
  {Radonji{\'{c}}}}, \bibinfo {author} {\bibfnamefont {Wassilij}\ \bibnamefont
  {Kopylov}}, \bibinfo {author} {\bibfnamefont {Antun}\ \bibnamefont
  {Bala{\v{z}}}}, \ and\ \bibinfo {author} {\bibfnamefont {Axel}\ \bibnamefont
  {Pelster}},\ }\bibfield  {title} {\enquote {\bibinfo {title} {Interplay of
  coherent and dissipative dynamics in condensates of light},}\ }\href
  {\doibase 10.1088/1367-2630/aac2a6} {\bibfield  {journal} {\bibinfo
  {journal} {New Journal of Physics}\ }\textbf {\bibinfo {volume} {20}},\
  \bibinfo {pages} {055014} (\bibinfo {year} {2018})}\BibitemShut {NoStop}%
\bibitem [{\citenamefont {Marthaler}\ \emph {et~al.}(2011)\citenamefont
  {Marthaler}, \citenamefont {Utsumi}, \citenamefont {Golubev}, \citenamefont
  {Shnirman},\ and\ \citenamefont {Sch\"on}}]{Marthaler2011}%
  \BibitemOpen
  \bibfield  {author} {\bibinfo {author} {\bibfnamefont {M.}~\bibnamefont
  {Marthaler}}, \bibinfo {author} {\bibfnamefont {Y.}~\bibnamefont {Utsumi}},
  \bibinfo {author} {\bibfnamefont {D.~S.}\ \bibnamefont {Golubev}}, \bibinfo
  {author} {\bibfnamefont {A.}~\bibnamefont {Shnirman}}, \ and\ \bibinfo
  {author} {\bibfnamefont {Gerd}\ \bibnamefont {Sch\"on}},\ }\bibfield  {title}
  {\enquote {\bibinfo {title} {Lasing without inversion in circuit quantum
  electrodynamics},}\ }\href {\doibase 10.1103/PhysRevLett.107.093901}
  {\bibfield  {journal} {\bibinfo  {journal} {Phys. Rev. Lett.}\ }\textbf
  {\bibinfo {volume} {107}},\ \bibinfo {pages} {093901} (\bibinfo {year}
  {2011})}\BibitemShut {NoStop}%
\bibitem [{\citenamefont {Lebreuilly}\ \emph {et~al.}(2018)\citenamefont
  {Lebreuilly}, \citenamefont {Chiocchetta},\ and\ \citenamefont
  {Carusotto}}]{lebreuilly2018pseudothermalization}%
  \BibitemOpen
  \bibfield  {author} {\bibinfo {author} {\bibfnamefont {Jos\'e}\ \bibnamefont
  {Lebreuilly}}, \bibinfo {author} {\bibfnamefont {Alessio}\ \bibnamefont
  {Chiocchetta}}, \ and\ \bibinfo {author} {\bibfnamefont {Iacopo}\
  \bibnamefont {Carusotto}},\ }\bibfield  {title} {\enquote {\bibinfo {title}
  {Pseudothermalization in driven-dissipative non-markovian open quantum
  systems},}\ }\href {\doibase 10.1103/PhysRevA.97.033603} {\bibfield
  {journal} {\bibinfo  {journal} {Phys. Rev. A}\ }\textbf {\bibinfo {volume}
  {97}},\ \bibinfo {pages} {033603} (\bibinfo {year} {2018})}\BibitemShut
  {NoStop}%
\bibitem [{\citenamefont {Mivehvar}\ \emph {et~al.}(2021)\citenamefont
  {Mivehvar}, \citenamefont {Piazza}, \citenamefont {Donner},\ and\
  \citenamefont {Ritsch}}]{Piazza2021}%
  \BibitemOpen
  \bibfield  {author} {\bibinfo {author} {\bibfnamefont {Farokh}\ \bibnamefont
  {Mivehvar}}, \bibinfo {author} {\bibfnamefont {Francesco}\ \bibnamefont
  {Piazza}}, \bibinfo {author} {\bibfnamefont {Tobias}\ \bibnamefont {Donner}},
  \ and\ \bibinfo {author} {\bibfnamefont {Helmut}\ \bibnamefont {Ritsch}},\
  }\bibfield  {title} {\enquote {\bibinfo {title} {Cavity qed with quantum
  gases: new paradigms in many-body physics},}\ }\href {\doibase
  10.1080/00018732.2021.1969727} {\bibfield  {journal} {\bibinfo  {journal}
  {Advances in Physics}\ }\textbf {\bibinfo {volume} {70}},\ \bibinfo {pages}
  {1--153} (\bibinfo {year} {2021})}\BibitemShut {NoStop}%
\bibitem [{\citenamefont {Fowler-Wright}\ \emph {et~al.}(2022)\citenamefont
  {Fowler-Wright}, \citenamefont {Lovett},\ and\ \citenamefont
  {Keeling}}]{Keeling_2022}%
  \BibitemOpen
  \bibfield  {author} {\bibinfo {author} {\bibfnamefont {Piper}\ \bibnamefont
  {Fowler-Wright}}, \bibinfo {author} {\bibfnamefont {Brendon~W.}\ \bibnamefont
  {Lovett}}, \ and\ \bibinfo {author} {\bibfnamefont {Jonathan}\ \bibnamefont
  {Keeling}},\ }\bibfield  {title} {\enquote {\bibinfo {title} {Efficient
  many-body non-markovian dynamics of organic polaritons},}\ }\href {\doibase
  10.1103/PhysRevLett.129.173001} {\bibfield  {journal} {\bibinfo  {journal}
  {Phys. Rev. Lett.}\ }\textbf {\bibinfo {volume} {129}},\ \bibinfo {pages}
  {173001} (\bibinfo {year} {2022})}\BibitemShut {NoStop}%
\bibitem [{\citenamefont {Arnardottir}\ \emph {et~al.}(2020)\citenamefont
  {Arnardottir}, \citenamefont {Moilanen}, \citenamefont {Strashko},
  \citenamefont {T\"orm\"a},\ and\ \citenamefont {Keeling}}]{Keeling_2020}%
  \BibitemOpen
  \bibfield  {author} {\bibinfo {author} {\bibfnamefont {Kristin~B.}\
  \bibnamefont {Arnardottir}}, \bibinfo {author} {\bibfnamefont {Antti~J.}\
  \bibnamefont {Moilanen}}, \bibinfo {author} {\bibfnamefont {Artem}\
  \bibnamefont {Strashko}}, \bibinfo {author} {\bibfnamefont {P\"aivi}\
  \bibnamefont {T\"orm\"a}}, \ and\ \bibinfo {author} {\bibfnamefont
  {Jonathan}\ \bibnamefont {Keeling}},\ }\bibfield  {title} {\enquote {\bibinfo
  {title} {Multimode organic polariton lasing},}\ }\href {\doibase
  10.1103/PhysRevLett.125.233603} {\bibfield  {journal} {\bibinfo  {journal}
  {Phys. Rev. Lett.}\ }\textbf {\bibinfo {volume} {125}},\ \bibinfo {pages}
  {233603} (\bibinfo {year} {2020})}\BibitemShut {NoStop}%
\bibitem [{\citenamefont {Fowler-Wright}\ \emph {et~al.}(2023)\citenamefont
  {Fowler-Wright}, \citenamefont {Arnard\'ottir}, \citenamefont {Kirton},
  \citenamefont {Lovett},\ and\ \citenamefont
  {Keeling}}]{PhysRevResearch.5.033148}%
  \BibitemOpen
  \bibfield  {author} {\bibinfo {author} {\bibfnamefont {Piper}\ \bibnamefont
  {Fowler-Wright}}, \bibinfo {author} {\bibfnamefont {Krist\'{\i}n~B.}\
  \bibnamefont {Arnard\'ottir}}, \bibinfo {author} {\bibfnamefont {Peter}\
  \bibnamefont {Kirton}}, \bibinfo {author} {\bibfnamefont {Brendon~W.}\
  \bibnamefont {Lovett}}, \ and\ \bibinfo {author} {\bibfnamefont {Jonathan}\
  \bibnamefont {Keeling}},\ }\bibfield  {title} {\enquote {\bibinfo {title}
  {Determining the validity of cumulant expansions for central spin models},}\
  }\href {\doibase 10.1103/PhysRevResearch.5.033148} {\bibfield  {journal}
  {\bibinfo  {journal} {Phys. Rev. Res.}\ }\textbf {\bibinfo {volume} {5}},\
  \bibinfo {pages} {033148} (\bibinfo {year} {2023})}\BibitemShut {NoStop}%
\bibitem [{\citenamefont {Abrikosov}(1965)}]{abrikosov1965electron}%
  \BibitemOpen
  \bibfield  {author} {\bibinfo {author} {\bibfnamefont {A.~A.}\ \bibnamefont
  {Abrikosov}},\ }\bibfield  {title} {\enquote {\bibinfo {title} {Electron
  scattering on magnetic impurities in metals and anomalous resistivity
  effects},}\ }\href {\doibase 10.1103/PhysicsPhysiqueFizika.2.5} {\bibfield
  {journal} {\bibinfo  {journal} {Physics Physique Fizika}\ }\textbf {\bibinfo
  {volume} {2}},\ \bibinfo {pages} {5--20} (\bibinfo {year}
  {1965})}\BibitemShut {NoStop}%
\bibitem [{\citenamefont {Barnes}(1976)}]{Barnes1977}%
  \BibitemOpen
  \bibfield  {author} {\bibinfo {author} {\bibfnamefont {S~E}\ \bibnamefont
  {Barnes}},\ }\bibfield  {title} {\enquote {\bibinfo {title} {{New method for
  the Anderson model}},}\ }\href {\doibase 10.1088/0305-4608/6/7/018}
  {\bibfield  {journal} {\bibinfo  {journal} {Journal of Physics F: Metal
  Physics}\ }\textbf {\bibinfo {volume} {6}},\ \bibinfo {pages} {1375}
  (\bibinfo {year} {1976})}\BibitemShut {NoStop}%
\bibitem [{\citenamefont {Barnes}(1977)}]{Barnes1977_II}%
  \BibitemOpen
  \bibfield  {author} {\bibinfo {author} {\bibfnamefont {S~E}\ \bibnamefont
  {Barnes}},\ }\bibfield  {title} {\enquote {\bibinfo {title} {{New method for
  the Anderson model. II. The U=0 limit}},}\ }\href {\doibase
  10.1088/0305-4608/7/12/022} {\bibfield  {journal} {\bibinfo  {journal}
  {Journal of Physics F: Metal Physics}\ }\textbf {\bibinfo {volume} {7}},\
  \bibinfo {pages} {2637} (\bibinfo {year} {1977})}\BibitemShut {NoStop}%
\bibitem [{\citenamefont {Bickers}(1987)}]{Bickers1987}%
  \BibitemOpen
  \bibfield  {author} {\bibinfo {author} {\bibfnamefont {N.~E.}\ \bibnamefont
  {Bickers}},\ }\bibfield  {title} {\enquote {\bibinfo {title} {Review of
  techniques in the large-$n$ expansion for dilute magnetic alloys},}\ }\href
  {\doibase 10.1103/RevModPhys.59.845} {\bibfield  {journal} {\bibinfo
  {journal} {Rev. Mod. Phys.}\ }\textbf {\bibinfo {volume} {59}},\ \bibinfo
  {pages} {845--939} (\bibinfo {year} {1987})}\BibitemShut {NoStop}%
\bibitem [{\citenamefont {Coleman}(1984)}]{Coleman1984}%
  \BibitemOpen
  \bibfield  {author} {\bibinfo {author} {\bibfnamefont {Piers}\ \bibnamefont
  {Coleman}},\ }\bibfield  {title} {\enquote {\bibinfo {title} {{New approach
  to the mixed-valence problem}},}\ }\href {\doibase 10.1103/PhysRevB.29.3035}
  {\bibfield  {journal} {\bibinfo  {journal} {Phys. Rev. B}\ }\textbf {\bibinfo
  {volume} {29}},\ \bibinfo {pages} {3035--3044} (\bibinfo {year}
  {1984})}\BibitemShut {NoStop}%
\bibitem [{\citenamefont {Kroha}\ and\ \citenamefont
  {W{\"o}lfle}(1998)}]{kroha1998fermi}%
  \BibitemOpen
  \bibfield  {author} {\bibinfo {author} {\bibfnamefont {Johann}\ \bibnamefont
  {Kroha}}\ and\ \bibinfo {author} {\bibfnamefont {Peter}\ \bibnamefont
  {W{\"o}lfle}},\ }\bibfield  {title} {\enquote {\bibinfo {title} {{Fermi and
  non-Fermi liquid behavior in quantum impurity systems: Conserving slave boson
  theory}},}\ }\href {\doibase 10.1007/BFb0107485} {\bibfield  {journal}
  {\bibinfo  {journal} {Acta Phys. Pol. B}\ }\textbf {\bibinfo {volume} {29}},\
  \bibinfo {pages} {3781} (\bibinfo {year} {1998})}\BibitemShut {NoStop}%
\bibitem [{\citenamefont {Kroha}\ and\ \citenamefont
  {W{\"o}lfle}(2005)}]{kroha2005JPSJ}%
  \BibitemOpen
  \bibfield  {author} {\bibinfo {author} {\bibfnamefont {Johann}\ \bibnamefont
  {Kroha}}\ and\ \bibinfo {author} {\bibfnamefont {Peter}\ \bibnamefont
  {W{\"o}lfle}},\ }\bibfield  {title} {\enquote {\bibinfo {title} {{Conserving
  Diagrammatic Approximations for Quantum Impurity Models: NCA and CTMA}},}\
  }\href {\doibase 10.1143/JPSJ.74.16} {\bibfield  {journal} {\bibinfo
  {journal} {J. Phys. Soc Jap.}\ }\textbf {\bibinfo {volume} {74}},\ \bibinfo
  {pages} {16} (\bibinfo {year} {2005})}\BibitemShut {NoStop}%
\bibitem [{\citenamefont {Anderson}(1961)}]{anderson1961localized}%
  \BibitemOpen
  \bibfield  {author} {\bibinfo {author} {\bibfnamefont {P.~W.}\ \bibnamefont
  {Anderson}},\ }\bibfield  {title} {\enquote {\bibinfo {title} {Localized
  magnetic states in metals},}\ }\href {\doibase 10.1103/PhysRev.124.41}
  {\bibfield  {journal} {\bibinfo  {journal} {Phys. Rev.}\ }\textbf {\bibinfo
  {volume} {124}},\ \bibinfo {pages} {41--53} (\bibinfo {year}
  {1961})}\BibitemShut {NoStop}%
\bibitem [{\citenamefont {Langreth}\ and\ \citenamefont
  {Nordlander}(1991)}]{Langreth1991}%
  \BibitemOpen
  \bibfield  {author} {\bibinfo {author} {\bibfnamefont {David~C.}\
  \bibnamefont {Langreth}}\ and\ \bibinfo {author} {\bibfnamefont
  {P.}~\bibnamefont {Nordlander}},\ }\bibfield  {title} {\enquote {\bibinfo
  {title} {{Derivation of a master equation for charge-transfer processes in
  atom-surface collisions}},}\ }\href {\doibase 10.1103/PhysRevB.43.2541}
  {\bibfield  {journal} {\bibinfo  {journal} {Phys. Rev. B}\ }\textbf {\bibinfo
  {volume} {43}},\ \bibinfo {pages} {2541--2557} (\bibinfo {year}
  {1991})}\BibitemShut {NoStop}%
\bibitem [{\citenamefont {Wingreen}\ and\ \citenamefont
  {Meir}(1994)}]{Wingreen1994}%
  \BibitemOpen
  \bibfield  {author} {\bibinfo {author} {\bibfnamefont {Ned~S.}\ \bibnamefont
  {Wingreen}}\ and\ \bibinfo {author} {\bibfnamefont {Yigal}\ \bibnamefont
  {Meir}},\ }\bibfield  {title} {\enquote {\bibinfo {title} {{Anderson model
  out of equilibrium: Noncrossing-approximation approach to transport through a
  quantum dot}},}\ }\href {\doibase 10.1103/PhysRevB.49.11040} {\bibfield
  {journal} {\bibinfo  {journal} {Phys. Rev. B}\ }\textbf {\bibinfo {volume}
  {49}},\ \bibinfo {pages} {11040--11052} (\bibinfo {year} {1994})}\BibitemShut
  {NoStop}%
\bibitem [{\citenamefont {Hettler}\ \emph {et~al.}(1998)\citenamefont
  {Hettler}, \citenamefont {Kroha},\ and\ \citenamefont
  {Hershfield}}]{hettler1998nonequilibrium}%
  \BibitemOpen
  \bibfield  {author} {\bibinfo {author} {\bibfnamefont {Matthias~H.}\
  \bibnamefont {Hettler}}, \bibinfo {author} {\bibfnamefont {Johann}\
  \bibnamefont {Kroha}}, \ and\ \bibinfo {author} {\bibfnamefont {Selman}\
  \bibnamefont {Hershfield}},\ }\bibfield  {title} {\enquote {\bibinfo {title}
  {{Nonequilibrium dynamics of the Anderson impurity model}},}\ }\href
  {\doibase 10.1103/PhysRevB.58.5649} {\bibfield  {journal} {\bibinfo
  {journal} {Phys. Rev. B}\ }\textbf {\bibinfo {volume} {58}},\ \bibinfo
  {pages} {5649--5664} (\bibinfo {year} {1998})}\BibitemShut {NoStop}%
\bibitem [{\citenamefont {Kroha}\ and\ \citenamefont
  {Zawadowski}(2002)}]{kroha2002nonequilibrium}%
  \BibitemOpen
  \bibfield  {author} {\bibinfo {author} {\bibfnamefont {J.}~\bibnamefont
  {Kroha}}\ and\ \bibinfo {author} {\bibfnamefont {A.}~\bibnamefont
  {Zawadowski}},\ }\bibfield  {title} {\enquote {\bibinfo {title}
  {{Nonequilibrium Quasiparticle Distribution Induced by Kondo Defects}},}\
  }\href {\doibase 10.1103/PhysRevLett.88.176803} {\bibfield  {journal}
  {\bibinfo  {journal} {Phys. Rev. Lett.}\ }\textbf {\bibinfo {volume} {88}},\
  \bibinfo {pages} {176803} (\bibinfo {year} {2002})}\BibitemShut {NoStop}%
\bibitem [{\citenamefont {{Eckstein, Martin and Werner,
  Philipp}}(2010)}]{Eckstein2010}%
  \BibitemOpen
  \bibfield  {author} {\bibinfo {author} {\bibnamefont {{Eckstein, Martin and
  Werner, Philipp}}},\ }\bibfield  {title} {\enquote {\bibinfo {title}
  {Nonequilibrium dynamical mean-field calculations based on the noncrossing
  approximation and its generalizations},}\ }\href {\doibase
  10.1103/PhysRevB.82.115115} {\bibfield  {journal} {\bibinfo  {journal} {Phys.
  Rev. B}\ }\textbf {\bibinfo {volume} {82}},\ \bibinfo {pages} {115115}
  (\bibinfo {year} {2010})}\BibitemShut {NoStop}%
\bibitem [{\citenamefont {Kamenev}(2023)}]{kamenev2011field}%
  \BibitemOpen
  \bibfield  {author} {\bibinfo {author} {\bibfnamefont {Alex}\ \bibnamefont
  {Kamenev}},\ }\href {\doibase 10.1017/CBO9781139003667} {\emph {\bibinfo
  {title} {{Field Theory of Non-Equilibrium Systems}}}}\ (\bibinfo  {publisher}
  {Cambridge University Press},\ \bibinfo {year} {2023})\BibitemShut {NoStop}%
\bibitem [{\citenamefont {Berges}(2004)}]{berges2004introduction}%
  \BibitemOpen
  \bibfield  {author} {\bibinfo {author} {\bibfnamefont {J{\"u}rgen}\
  \bibnamefont {Berges}},\ }\bibfield  {title} {\enquote {\bibinfo {title}
  {{Introduction to Nonequilibrium Quantum Field Theory}},}\ }\href@noop {}
  {\bibfield  {journal} {\bibinfo  {journal} {AIP Conference Proceedings}\
  }\textbf {\bibinfo {volume} {739}},\ \bibinfo {pages} {3--62} (\bibinfo
  {year} {2004})}\BibitemShut {NoStop}%
\bibitem [{\citenamefont {Gasenzer}\ \emph {et~al.}(2005)\citenamefont
  {Gasenzer}, \citenamefont {Berges}, \citenamefont {Schmidt},\ and\
  \citenamefont {Seco}}]{Gasenzer2005}%
  \BibitemOpen
  \bibfield  {author} {\bibinfo {author} {\bibfnamefont {Thomas}\ \bibnamefont
  {Gasenzer}}, \bibinfo {author} {\bibfnamefont {J\"urgen}\ \bibnamefont
  {Berges}}, \bibinfo {author} {\bibfnamefont {Michael~G.}\ \bibnamefont
  {Schmidt}}, \ and\ \bibinfo {author} {\bibfnamefont {Marcos}\ \bibnamefont
  {Seco}},\ }\bibfield  {title} {\enquote {\bibinfo {title} {Nonperturbative
  dynamical many-body theory of a bose-einstein condensate},}\ }\href {\doibase
  10.1103/PhysRevA.72.063604} {\bibfield  {journal} {\bibinfo  {journal} {Phys.
  Rev. A}\ }\textbf {\bibinfo {volume} {72}},\ \bibinfo {pages} {063604}
  (\bibinfo {year} {2005})}\BibitemShut {NoStop}%
\bibitem [{\citenamefont {Bock}\ \emph {et~al.}(2016)\citenamefont {Bock},
  \citenamefont {Liluashvili},\ and\ \citenamefont {Gasenzer}}]{Bock2016}%
  \BibitemOpen
  \bibfield  {author} {\bibinfo {author} {\bibfnamefont {Sebastian}\
  \bibnamefont {Bock}}, \bibinfo {author} {\bibfnamefont {Alexander}\
  \bibnamefont {Liluashvili}}, \ and\ \bibinfo {author} {\bibfnamefont
  {Thomas}\ \bibnamefont {Gasenzer}},\ }\bibfield  {title} {\enquote {\bibinfo
  {title} {{Buildup of the Kondo effect from real-time effective action for the
  Anderson impurity model}},}\ }\href {\doibase 10.1103/PhysRevB.94.045108}
  {\bibfield  {journal} {\bibinfo  {journal} {Phys. Rev. B}\ }\textbf {\bibinfo
  {volume} {94}},\ \bibinfo {pages} {045108} (\bibinfo {year}
  {2016})}\BibitemShut {NoStop}%
\bibitem [{\citenamefont {Stan}\ \emph {et~al.}(2009)\citenamefont {Stan},
  \citenamefont {Dahlen},\ and\ \citenamefont {van Leeuwen}}]{Stan2009}%
  \BibitemOpen
  \bibfield  {author} {\bibinfo {author} {\bibfnamefont {Adrian}\ \bibnamefont
  {Stan}}, \bibinfo {author} {\bibfnamefont {Nils~Erik}\ \bibnamefont
  {Dahlen}}, \ and\ \bibinfo {author} {\bibfnamefont {Robert}\ \bibnamefont
  {van Leeuwen}},\ }\bibfield  {title} {\enquote {\bibinfo {title} {Time
  propagation of the kadanoff–baym equations for inhomogeneous systems},}\
  }\href {\doibase 10.1063/1.3127247} {\bibfield  {journal} {\bibinfo
  {journal} {The Journal of Chemical Physics}\ }\textbf {\bibinfo {volume}
  {130}},\ \bibinfo {pages} {224101} (\bibinfo {year} {2009})}\BibitemShut
  {NoStop}%
\bibitem [{\citenamefont {Meirinhos}\ \emph {et~al.}(2022)\citenamefont
  {Meirinhos}, \citenamefont {Kajan}, \citenamefont {Kroha},\ and\
  \citenamefont {Bode}}]{Meirinhos2022}%
  \BibitemOpen
  \bibfield  {author} {\bibinfo {author} {\bibfnamefont {Francisco}\
  \bibnamefont {Meirinhos}}, \bibinfo {author} {\bibfnamefont {Michael}\
  \bibnamefont {Kajan}}, \bibinfo {author} {\bibfnamefont {Johann}\
  \bibnamefont {Kroha}}, \ and\ \bibinfo {author} {\bibfnamefont {Tim}\
  \bibnamefont {Bode}},\ }\bibfield  {title} {\enquote {\bibinfo {title}
  {{Adaptive numerical solution of Kadanoff-Baym equations}},}\ }\href
  {\doibase 10.21468/SciPostPhysCore.5.2.030} {\bibfield  {journal} {\bibinfo
  {journal} {SciPost Phys. Core}\ }\textbf {\bibinfo {volume} {5}},\ \bibinfo
  {pages} {030} (\bibinfo {year} {2022})}\BibitemShut {NoStop}%
\bibitem [{\citenamefont {Carmichael}(2013)}]{carmichael2013statistical}%
  \BibitemOpen
  \bibfield  {author} {\bibinfo {author} {\bibfnamefont {Howard~J}\
  \bibnamefont {Carmichael}},\ }\href {\doibase 10.1007/978-3-662-03875-8}
  {\emph {\bibinfo {title} {{Statistical Methods in Quantum Optics 1: Master
  Equations and Fokker-Planck Equations}}}}\ (\bibinfo  {publisher} {Springer
  Berlin, Heidelberg},\ \bibinfo {year} {2013})\BibitemShut {NoStop}%
\bibitem [{\citenamefont {Heiss}(2012)}]{heiss2012physics}%
  \BibitemOpen
  \bibfield  {author} {\bibinfo {author} {\bibfnamefont {W~D}\ \bibnamefont
  {Heiss}},\ }\bibfield  {title} {\enquote {\bibinfo {title} {The physics of
  exceptional points},}\ }\href {\doibase 10.1088/1751-8113/45/44/444016}
  {\bibfield  {journal} {\bibinfo  {journal} {Journal of Physics A:
  Mathematical and Theoretical}\ }\textbf {\bibinfo {volume} {45}},\ \bibinfo
  {pages} {444016} (\bibinfo {year} {2012})}\BibitemShut {NoStop}%
\bibitem [{\citenamefont {Kasprzak}\ \emph {et~al.}(2006)\citenamefont
  {Kasprzak}, \citenamefont {Richard}, \citenamefont {Kundermann},
  \citenamefont {Baas}, \citenamefont {Jeambrun}, \citenamefont {Keeling},
  \citenamefont {Marchetti}, \citenamefont {Szyma{\'{n}}ska}, \citenamefont
  {Andr{\'e}}, \citenamefont {Staehli}, \citenamefont {Savona}, \citenamefont
  {Littlewood}, \citenamefont {Deveaud},\ and\ \citenamefont
  {Dang}}]{Littlewood2006}%
  \BibitemOpen
  \bibfield  {author} {\bibinfo {author} {\bibfnamefont {J.}~\bibnamefont
  {Kasprzak}}, \bibinfo {author} {\bibfnamefont {M.}~\bibnamefont {Richard}},
  \bibinfo {author} {\bibfnamefont {S.}~\bibnamefont {Kundermann}}, \bibinfo
  {author} {\bibfnamefont {A.}~\bibnamefont {Baas}}, \bibinfo {author}
  {\bibfnamefont {P.}~\bibnamefont {Jeambrun}}, \bibinfo {author}
  {\bibfnamefont {J.~M.~J.}\ \bibnamefont {Keeling}}, \bibinfo {author}
  {\bibfnamefont {F.~M.}\ \bibnamefont {Marchetti}}, \bibinfo {author}
  {\bibfnamefont {M.~H.}\ \bibnamefont {Szyma{\'{n}}ska}}, \bibinfo {author}
  {\bibfnamefont {R.}~\bibnamefont {Andr{\'e}}}, \bibinfo {author}
  {\bibfnamefont {J.~L.}\ \bibnamefont {Staehli}}, \bibinfo {author}
  {\bibfnamefont {V.}~\bibnamefont {Savona}}, \bibinfo {author} {\bibfnamefont
  {P.~B.}\ \bibnamefont {Littlewood}}, \bibinfo {author} {\bibfnamefont
  {B.}~\bibnamefont {Deveaud}}, \ and\ \bibinfo {author} {\bibfnamefont
  {Le~Si}\ \bibnamefont {Dang}},\ }\bibfield  {title} {\enquote {\bibinfo
  {title} {Bose--einstein condensation of exciton polaritons},}\ }\href
  {\doibase 10.1038/nature05131} {\bibfield  {journal} {\bibinfo  {journal}
  {Nature}\ }\textbf {\bibinfo {volume} {443}},\ \bibinfo {pages} {409--414}
  (\bibinfo {year} {2006})}\BibitemShut {NoStop}%
\bibitem [{\citenamefont {White}\ \emph {et~al.}(2020)\citenamefont {White},
  \citenamefont {Hill}, \citenamefont {Pollock}, \citenamefont {Hollenberg},\
  and\ \citenamefont {Modi}}]{White2020}%
  \BibitemOpen
  \bibfield  {author} {\bibinfo {author} {\bibfnamefont {G.~A.~L.}\
  \bibnamefont {White}}, \bibinfo {author} {\bibfnamefont {C.~D.}\ \bibnamefont
  {Hill}}, \bibinfo {author} {\bibfnamefont {F.~A.}\ \bibnamefont {Pollock}},
  \bibinfo {author} {\bibfnamefont {L.~C.~L.}\ \bibnamefont {Hollenberg}}, \
  and\ \bibinfo {author} {\bibfnamefont {K.}~\bibnamefont {Modi}},\ }\bibfield
  {title} {\enquote {\bibinfo {title} {Demonstration of non-markovian process
  characterisation and control on a quantum processor},}\ }\href {\doibase
  10.1038/s41467-020-20113-3} {\bibfield  {journal} {\bibinfo  {journal}
  {Nature Communications}\ }\textbf {\bibinfo {volume} {11}},\ \bibinfo {pages}
  {6301} (\bibinfo {year} {2020})}\BibitemShut {NoStop}%
\bibitem [{\citenamefont {Schwinger}(1961)}]{Schwinger1961}%
  \BibitemOpen
  \bibfield  {author} {\bibinfo {author} {\bibfnamefont {Julian}\ \bibnamefont
  {Schwinger}},\ }\bibfield  {title} {\enquote {\bibinfo {title} {Brownian
  motion of a quantum oscillator},}\ }\href {\doibase 10.1063/1.1703727}
  {\bibfield  {journal} {\bibinfo  {journal} {Journal of Mathematical Physics}\
  }\textbf {\bibinfo {volume} {2}},\ \bibinfo {pages} {407--432} (\bibinfo
  {year} {1961})}\BibitemShut {NoStop}%
\end{thebibliography}%

\end{document}